\documentclass[a4paper,11pt]{article}

\usepackage{jinstpub} 
\usepackage{graphicx}
\usepackage{xcolor}
\usepackage{caption}
\usepackage{subcaption}
\usepackage{multirow}
\usepackage{amsmath,amssymb}
\usepackage[numbers,sort&compress]{natbib}
\usepackage{threeparttable}
\usepackage{textcomp}

\title{Study of 20-inch PMTs dark count generated large pulses}

\author[a,b,c]{Yu Zhang,}
\author[a,b,c,1]{Zhimn Wang,\note{Corresponding author.}}
\author[a,b,c]{Min Li,}
\author[a,b,c]{Yongpeng Zhang,}
\author[a]{Yaoguang Wang,}
\author[a,b,c]{Zhaoyuan Peng,}
\author[a,b]{Changgen Yang,}
\author[a,b,c]{Yuekun Heng}

\affiliation[a]{Institute of High Energy Physics, Beijing 100049, China}
\affiliation[b]{University of Chinese Academy of Sciences, Beijing 100049, China}
\affiliation[c]{State Key Laboratory of Particle Detection and Electronics, Beijing 100049, China}

\emailAdd{wangzhm@ihep.ac.cn}

\abstract{The main goal of the JUNO experiment is to determine the neutrino mass ordering with a 20\,kt liquid-scintillator detector. The 20-inch PMT is crucial as one of JUNO key instruments to realize an excellent energy resolution of at least 3\,\% at 1\,MeV. The knowledge on PMT's characterisation and feature is critical for detector performance understanding. Large pulses from PMT dark count such as from flasher or others are one of the serious concerns for detector noise control. Focusing on the large pulses from 20-inch PMT dark count, this paper is trying to investigate the causes by measurements with a muon tagging system. It is found that the large pules of 20-inch PMT dark count is contributed mainly from muons hitting the PMT glass. A simulation is also realized and achieved a consistent understanding.}


\arxivnumber{1234.56789} 

\date{Received: date / Accepted: date}

\begin{document}
\maketitle
\flushbottom

\section{Introduction}
\label{1:intro}

\begin{figure*}[!htb]
	\centering
	\begin{subfigure}[c]{0.4\textwidth}
	\centering
	\includegraphics[width=\linewidth]{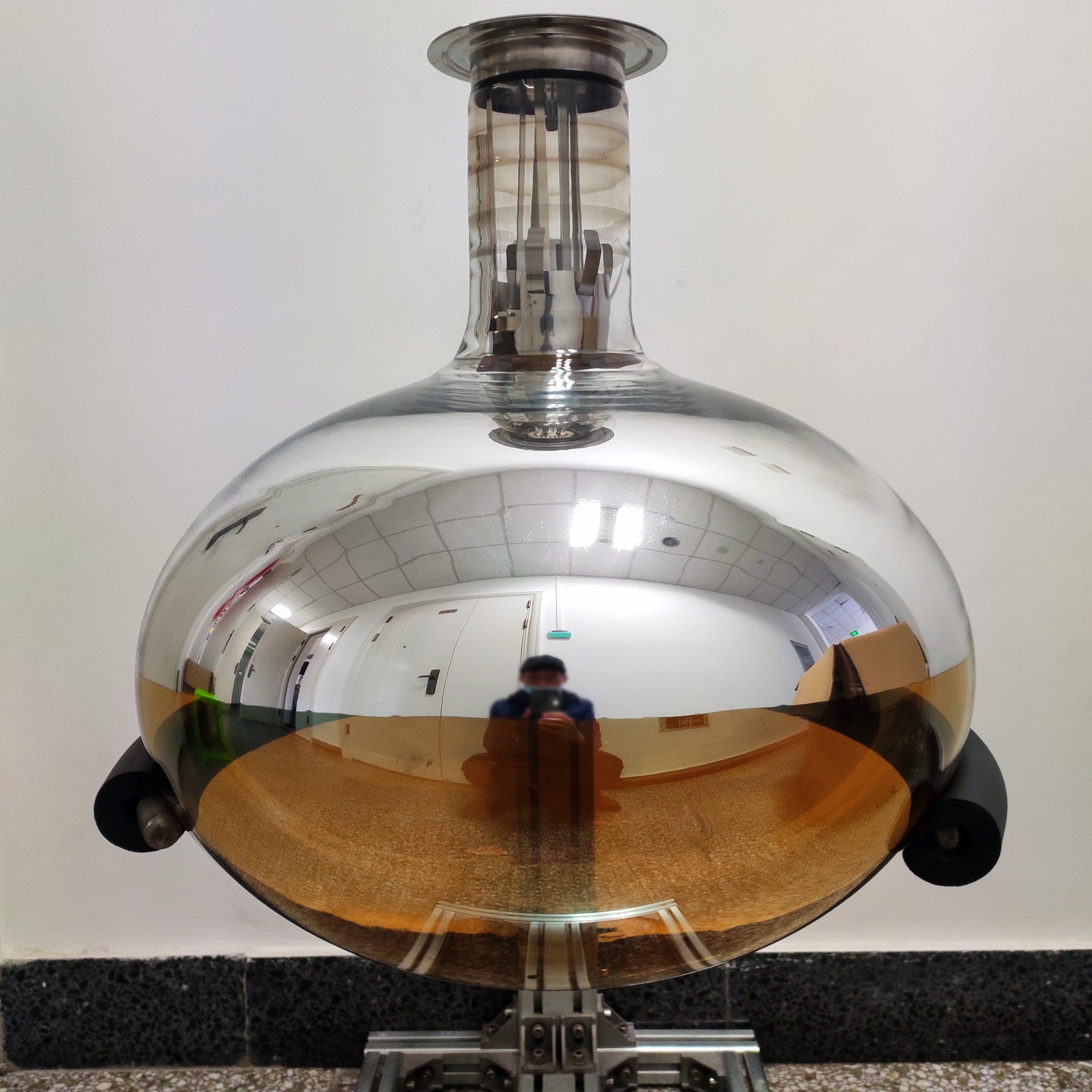}
	\end{subfigure}
	\begin{subfigure}[c]{0.4\textwidth}
	\centering
	\includegraphics[width=\linewidth]{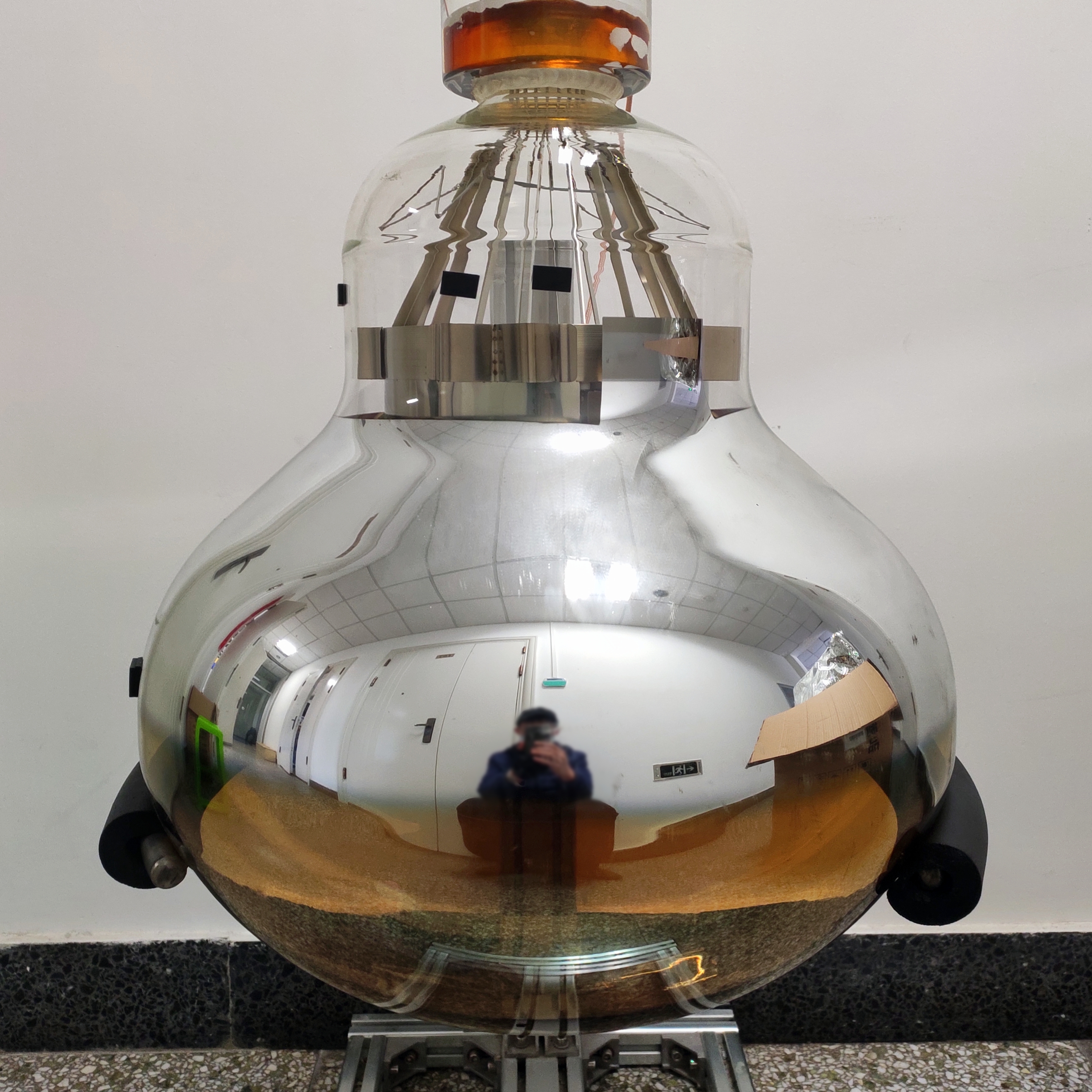}
	\end{subfigure}	
    \caption{Photograph of JUNO selected 20-inch NNVT MCP PMT (left) and HPK dynode PMT (right).}
    \label{fig:PMT}
\end{figure*}

The Jiangmen Underground Neutrino Observatory (JUNO)\,\cite{JUNOCDR,JUNOphysics} is under construction at Jiangmen, Guangdong, China. The experiment aims to study neutrino mass ordering with 3\% energy resolution at 1\,MeV, a precise determination of neutrino oscillation parameters, and other neutrino physics with 20\,kton liquid scintillator viewed by up to 20,000 high quantum efficiency (QE) 20-inch PMTs. JUNO selected two kinds of 20-inch PMTs whose typical photon detection efficiency higher than 27\%\,\cite{JUNOdetector,JUNOPMTliangjian,JUNOPMTinstr}: 5,000 of Hamamatsu Photonics K.K. (HPK, R12860) dynode PMT\,\cite{HPK-R12860} and 15,000 of the newly developed MCP PMT from North Night Vision Technology Co., LTD (NNVT, GDB6201)\,\cite{NNVT-GDB6201-note}, as shown in Fig.\,\ref{fig:PMT}.

Photomultiplier tubes (PMT) are widely used in particle physics experiments for the light detection sensitive to single photon, such as Super-K\,\cite{Super-Kamiokande:1998uiq,PhysRevD.83.052010}, KamLAND\,\cite{PhysRevLett.90.021802}, SNO\,\cite{SNO}, MiniBoone\,\cite{MiniBoone}, Icecube\,\cite{PhysRevLett.110.131302}, Chooz\,\cite{chooz}, Daya Bay\,\cite{dayabay}, and RENO\,\cite{KIM201324}. PMT‘s performance and its characters have been studied in detail with well understanding\,\cite{AugerPMT,GE2016175,BorexinoPMT,DayabayPMT,ChoozPMT,HKPMT,JUNO3inchPMT,JUNOPMTinstr,KM3NeTPMT,JUNOPMTflasher,MCPPMT2018,YWang_newMCP,wavesamplingPMT,waveAnalysisHaiqiong}. Some large pulses of PMT made a lot of trouble to recent world-wide rare event neutrino experiments, and additional analysis strategy is introduced to suppress its effect, such as Double Chooz\cite{Abe_2016}, Daya Bay\cite{DWYER201330}, Icecube\cite{IceCube-inproceedings} and RENO\cite{JANG2014145}. Additional studies are done on flasher and large pulses already as in\,\cite{Yang_2020,Qian_2020}. It is also valuable to mention that the flashers from PMT HV divider of JUNO is studied in \cite{JUNOPMTflasher}. As known, it will generate a large pulse by Cerenkov radiation when a muon is crossing through the PMT glass (Hamamatsu PMT R5912) as studied in\,\cite{PMTmuon2007,BAYAT20141}. PMT large pulse still needs further study to identify either flasher or others to have a better rejection to noise, in particular for the 20-inch PMTs and the coming rare event detection projects, such as JUNO\cite{JUNOCDR,JUNOphysics}, HyperK\cite{Hyper-Kamiokande:2018ofw,HK10.1093/ptep/ptv061}.

In this article, we will show a detailed study on the large pulses of 20-inch PMT associated with muon tagged by a plastic scintillator coincidence system. Sec.\,\ref{1:setup} will provide a short description on the testing system and configurations. Sec.\,\ref{1:results} will show the measurement results under different conditions. A dedicated simulation is done for better understanding and comparing with the measurements as shown in Sec.\,\ref{1:sim}. Finally, a short summary is reached in Sec.\,\ref{1:summary}.

\section{Experimental setup}
\label{1:setup}

\begin{figure}[!htb]
	\begin{subfigure}[c]{0.415\textwidth}
	\centering
		\includegraphics[width=\linewidth]{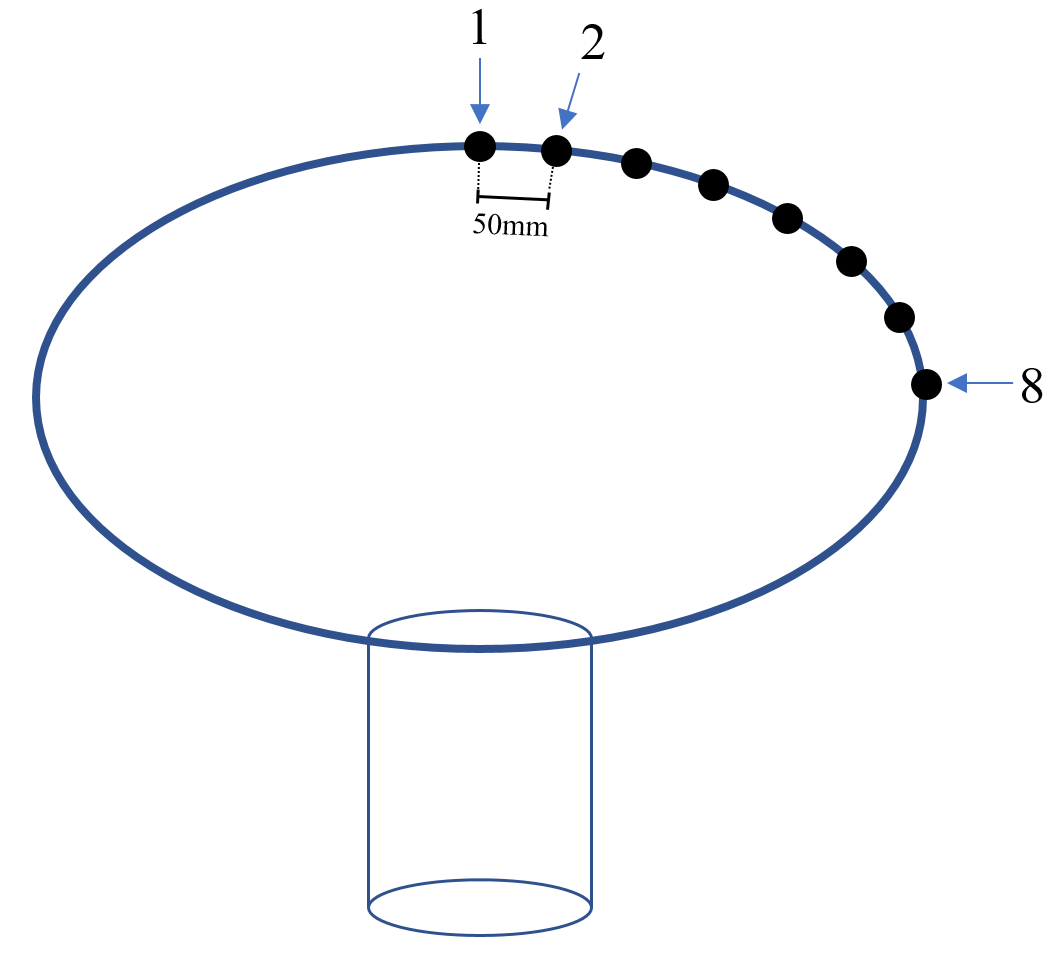}
	\end{subfigure}\hfill
	\begin{subfigure}[c]{0.56\textwidth}
	\centering
	\includegraphics[width=\linewidth]{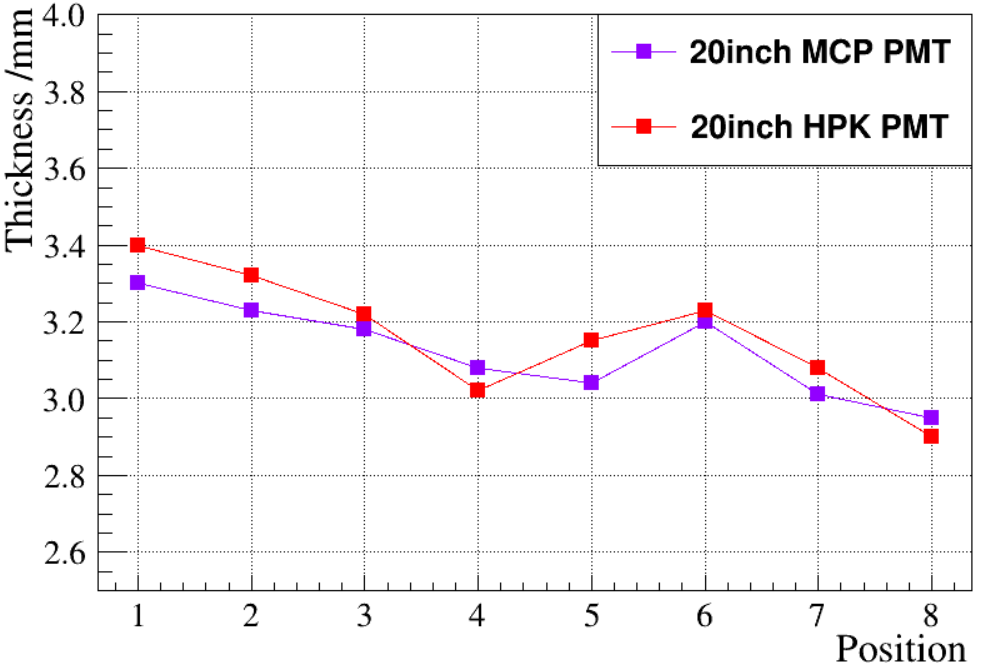}
	\end{subfigure}	
	\caption{Glass thickness of the photocathode hemisphere of 20-inch PMTs. Left: measurement positions at different zenith angles; right: the measured glass thickness of HPK and NNVT PMTs.}
	\label{fig:PMT:thickness}       
\end{figure}

For better pressure tolerance of both types of JUNO selected PMTs, the thickness is designed larger for most part of the glass bulb. The typical glass thickness\footnote{The thickness of 20-inch NNVT PMT and HPK PMT glass at different positions is measured by an ultrasonic thickness gauge TT112\cite{TT112}.} is measured with few PMT samples as shown in Fig.\,\ref{fig:PMT:thickness}, which is thicker than the widely used Hamamatsu R5912 1-3\,mm\,\cite{R5912-CHOW201525}. Considering the huge dimension, high QE and larger thickness of the glass bulb of the JUNO selected 20-inch PMTs, it is expected to have a higher muon hitting rate at sea level and stronger Cerenkov light intensity for a muon passing through the glass.

\subsection{Testing System}
\label{2:system}
\begin{figure*}[!htb]
    \centering
    \includegraphics[scale=0.18]{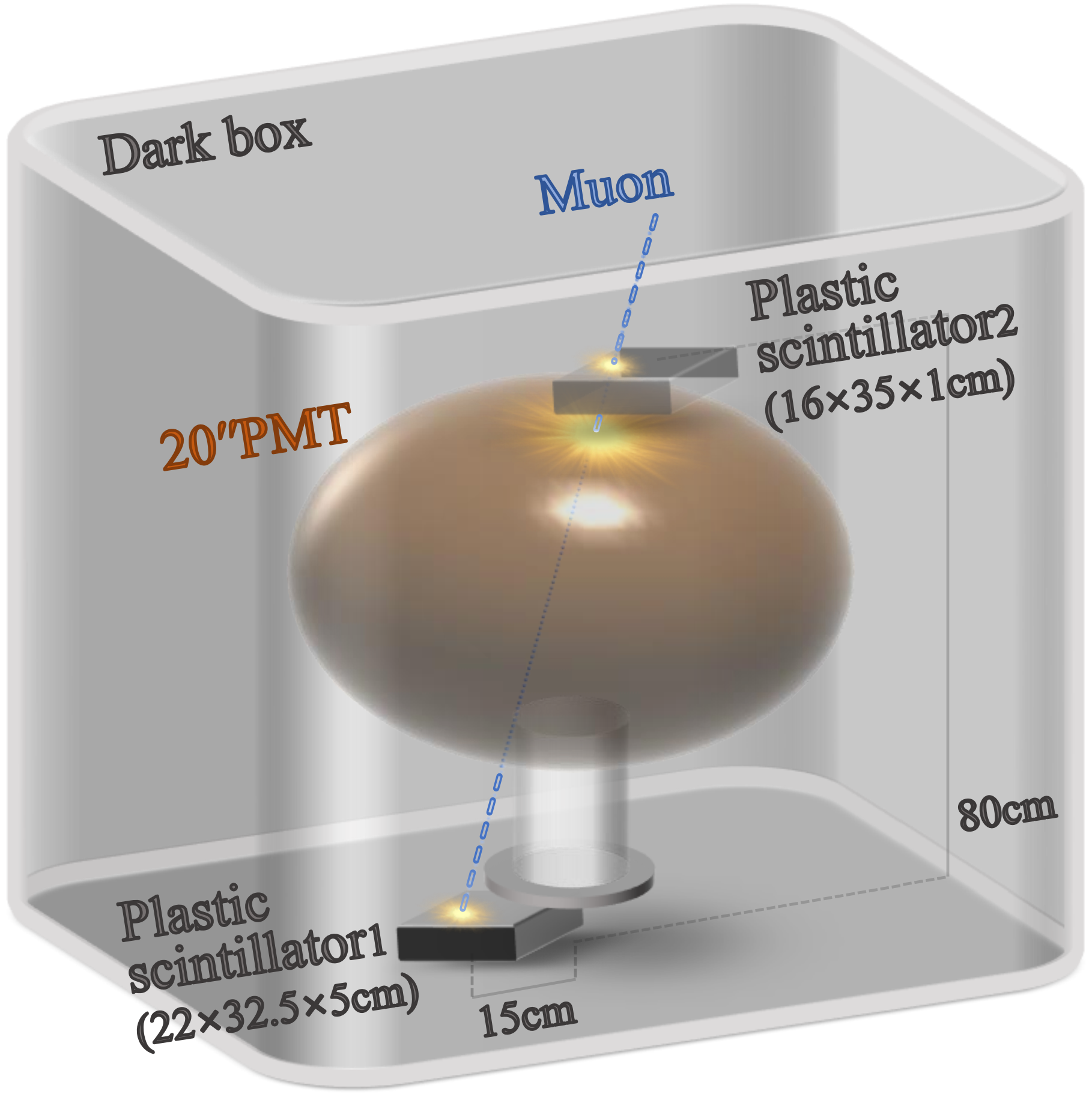}
    \caption{Layout of the 20-inch PMT's large pulse testing system with muon tracking scintillators.}
    \label{fig:testing:schema}
\end{figure*}

\begin{figure*}[!htb]
    \centering
    \includegraphics[scale=0.3]{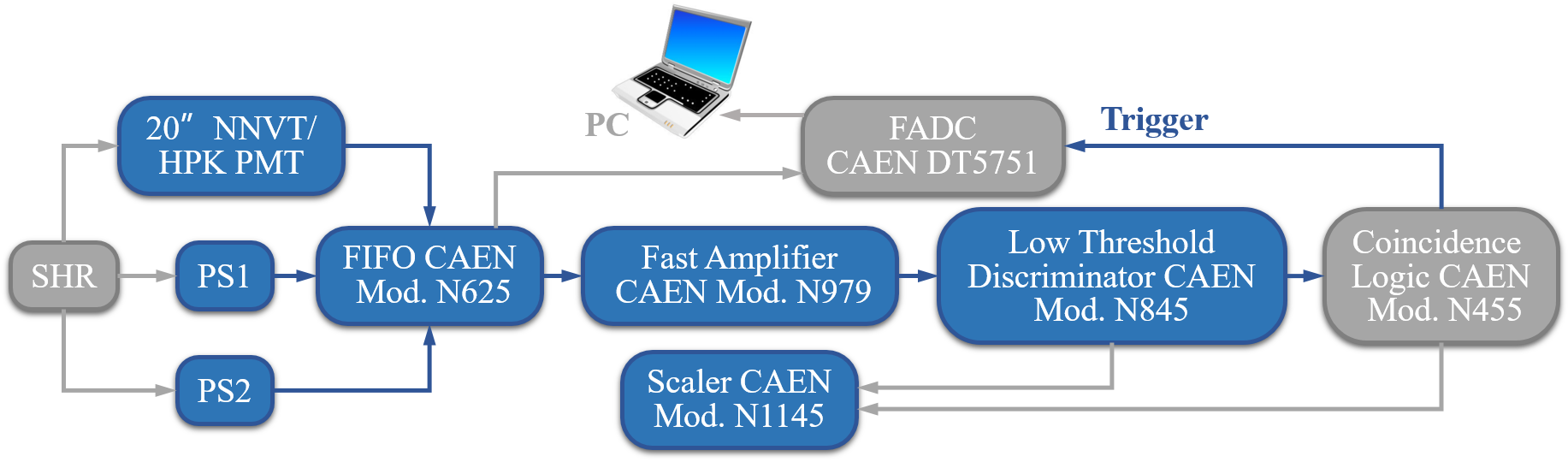}
    \caption{Electronics system used for the large pulse testing of 20-inch PMT dark count.}
    \label{fig:signal:processing}
\end{figure*}

A muon coincidence system is used to identify the 20” PMT's pulse correlated with muons from different directions by two plastic scintillators as shown in Fig.\,\ref{fig:testing:schema}. A muon will be tagged by the coincidence of the two plastic scintillators (PS1 and PS2, distance in height is around 80\,cm) to select the muons going through the glass bulb of the 20-inch PMT. The dimension of the scintillators is 22$\times$32.5$\times$5\,cm$^3$  for PS1, and 16$\times$35$\times$1\,cm$^3$ for PS2, respectively. PMT XP2020\,\cite{XP2020}, coupled to the scintillators by light guide, has a very narrow width of its output pulse (FWHM $\sim$2.4\,ns), which is much smaller than the 20-inch PMTs (FWHM 7$\sim$10\,ns). The PS1 will be moved to several locations to select muons with different incident angles, and the 20-inch PMT can be placed with glass bulb up (as shown in Fig.\,\ref{fig:testing:schema}) or glass bulb down to check the effect of the Cerenkov light direction. The waveforms of the three signal channels from the two sinctillators and the 20-inch PMT will be acquired in a same time window following the coincidence trigger by a digitizer of CAEN DT5751\cite{CAEN-DT5751}, which has 1\,GSample/s, 1\,Vpp dynamic range with 1\,mV precision under 50\,$\Omega$ impedance. The power of all the used PMTs is supplied by a SHR desktop HV module\cite{iseg-SHR}. More detailed information on the signals' processing electronics is shown in Fig.\,\ref{fig:signal:processing}, where few electronics modules are used for signal splitting (Linear FIFO CAEN Mod. N625\cite{CAEN-N625}), $\times$10 fast amplification (CAEN Mod. N979\cite{CAEN-N979}), low threshold discrimination (CAEN Mod. N845\cite{CAEN-N845}), logic counter (CAEN Mod. N1145\cite{CAEN-N1145}), and logic coincidence (CAEN Mod. N455\cite{CAEN-N455}).

\begin{figure}[!htb]
    \centering
    \includegraphics[scale=0.2]{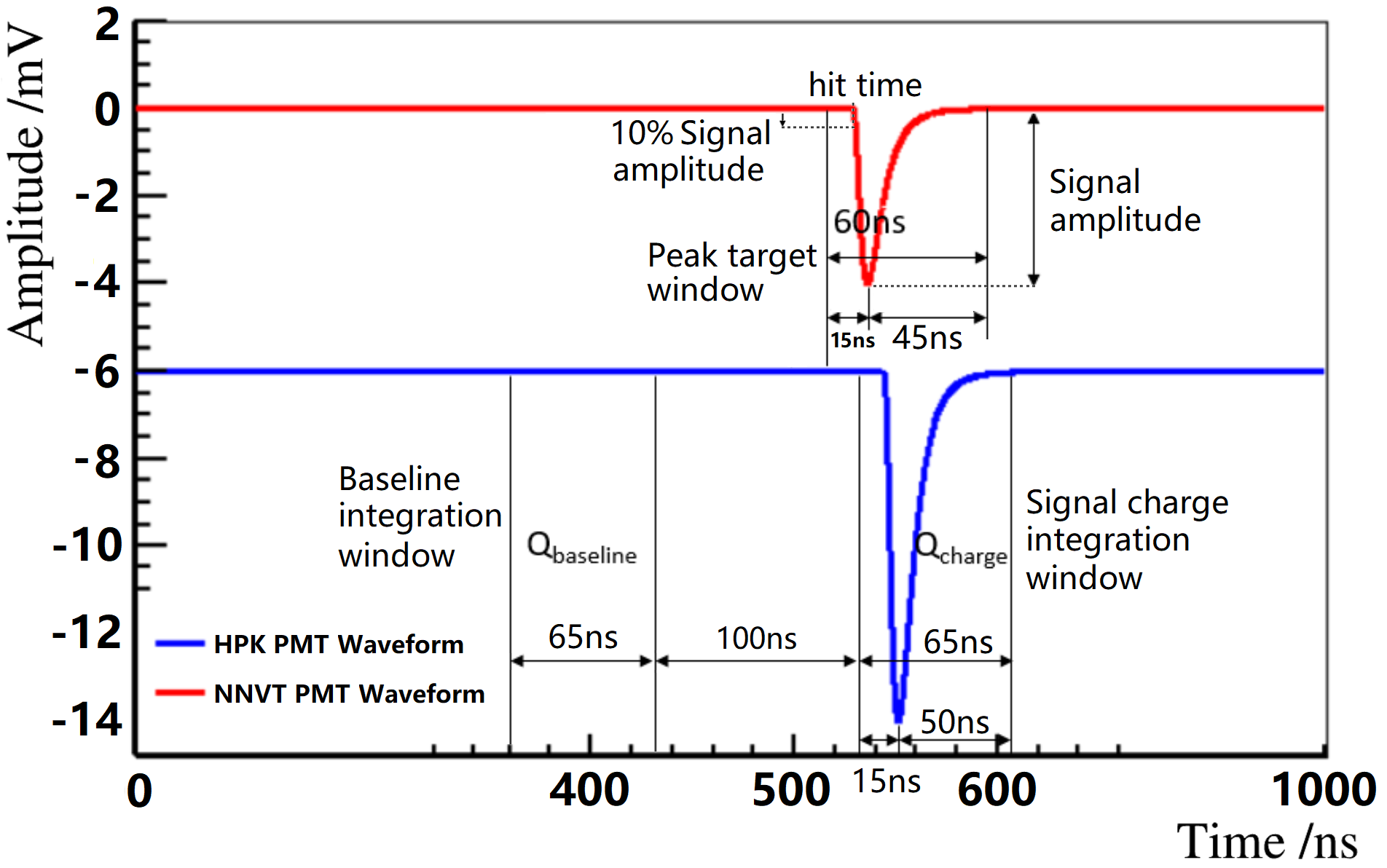}
    \caption{Exemplary of signal waveforms (simulation, with artificial offset): the charge integration window for baseline and signal, hit time, and pulse amplitude. Red: NNVT PMT; blue, HPK PMT.}
    \label{fig:pulse:integration}
\end{figure}

The coincidence window length used for trigger generation is set to 100\,ns, and the window length of the waveform data taking is selected to 1\,\textmu s, where the primary signal is shifted to around 450-650\,ns for both scintillators and 20-inch PMT. Furthermore, the system can be set to several trigger configurations manually (trigger mode):
\begin{itemize}
	\item[1.] Triggered by only one of the three signal channels (PMTs) for the measurement of dark count, mainly for the 20-inch PMT.
	\item[2.] Triggered by two-coincidence of PS1 and 20-inch PMT for muon and gammas which passes the 20-inch PMT.
	\item[3.] Triggered by triple-coincidence of PS1, PS2 and 20-inch PMT for purer muon which passes 20-inch PMT.
\end{itemize}
With the record waveforms, a charge integration window of 20-inch PMT is selected relative to the peak location of the primary pulse in [-15,45]\,ns for NNVT PMT and [-15,50]\,ns for HPK PMT, while the window for baseline is shifted before the primary pulse by 100\,ns and selected in [-115,-55]\,ns and [-115,-50]\,ns respectively as shown in Fig.\,\ref{fig:pulse:integration}. The hit time, pulse amplitude will be derived from the waveform too. The signal of plastic scintillators is processed similar to the 20-inch PMT.

\subsection{PMT settings}
\label{2:config}

The 20-inch PMT is working with an optimized HV divider and positive HV \cite{JUNOPMTsignalopt,JUNOPMTsignalover}.
The working point of the 20-inch PMT is aiming to be around a gain of $1\times 10^{7}$ for both types of PMTs to simulate the future JUNO conditions. Following the traditional PMT gain calibration methods, such as a DC current-based method\,\cite{POLYAKOV201369}, charge spectrum method with pulse light source\,\cite{PMTgainmodel1994,Luo_2019}, and gain determination algorithms\,\cite{PMTgainmodel2017,JUNOPMTgain}, a method (peak gain) based on the peak of single photoelectron (SPE) spectrum from dark count events is used to simply the testing and analysis process.

\begin{figure}[!htb]
	\begin{subfigure}[c]{0.495\textwidth}
	\centering
		\includegraphics[width=\linewidth]{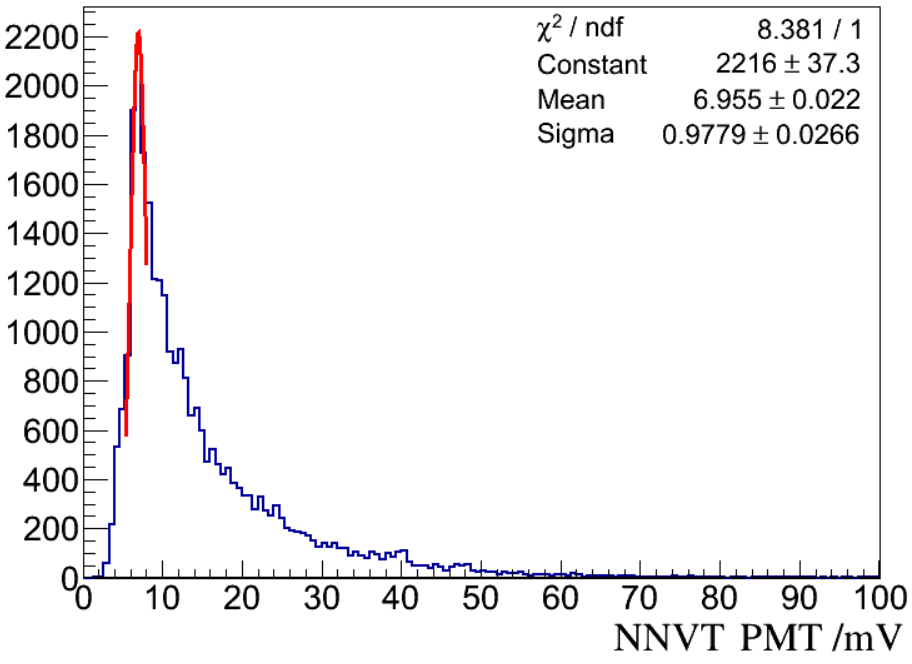}
	\end{subfigure}\hfill
	\begin{subfigure}[c]{0.495\textwidth}
	\centering
	\includegraphics[width=\linewidth]{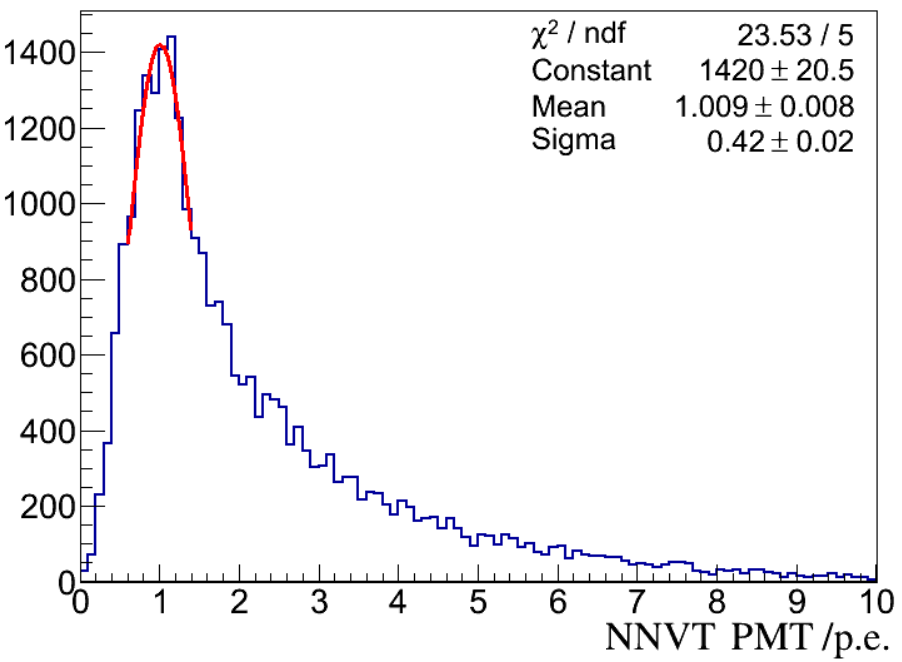}
	\end{subfigure}	
	\begin{subfigure}[c]{0.495\textwidth}
	\centering
		\includegraphics[width=\linewidth]{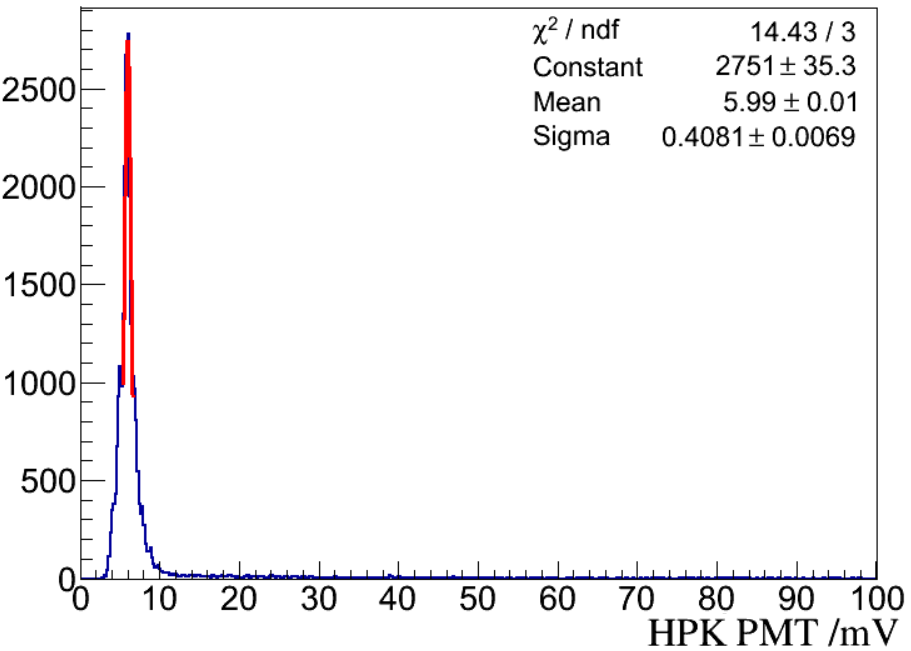}
	\end{subfigure}\hfill
	\begin{subfigure}[c]{0.495\textwidth}
	\centering
	\includegraphics[width=\linewidth]{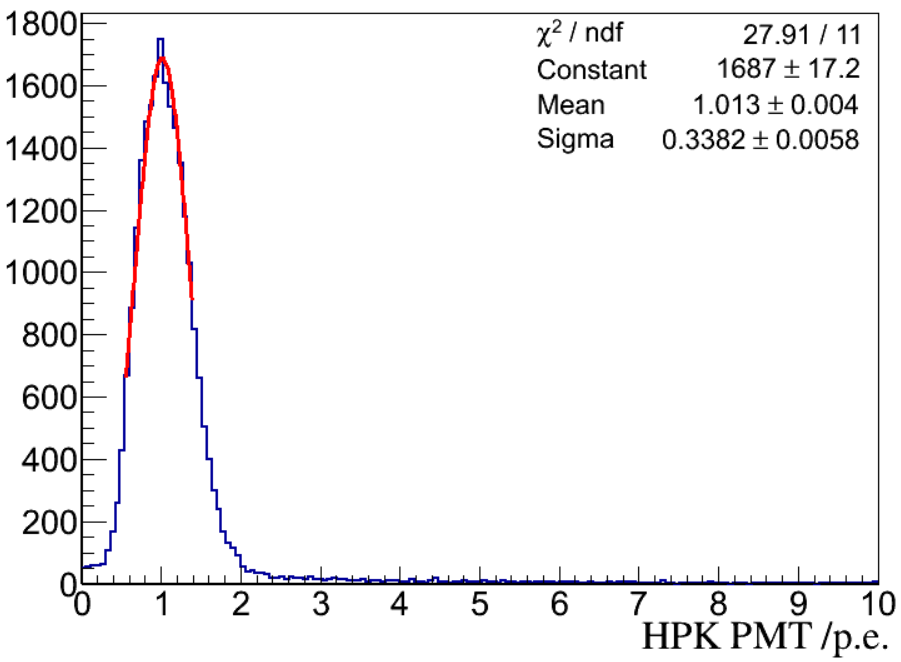}
	\end{subfigure}	
	\caption{Measured SPE spectra of 20-inch HPK or NNVT PMTs, where a Gaussian fit in red line is applied around the peaks. Top left: MCP PMT SPE amplitude spectrum; top right: MCP PMT SPE charge spectrum; bottom left: HPK PMT SPE amplitude spectrum; bottom right: HPK PMT SPE charge spectrum. Note that the calibrated PMT gain is used for the charge calculation in p.e..}
	\label{fig:PMT:spe}       
\end{figure}

Derived from the measured SPE spectra of charge and amplitude of 20-inch PMTs as shown in Fig.\,\ref{fig:PMT:spe}, the SPE amplitude is around 7.0\,mV for NNVT PMT with a gain of 1.2$\times 10^{7}$ and 6.0\,mV for HPK PMT with a gain of 1.0$\times 10^{7}$. An analysis threshold in amplitude of 3\,mV is used for both types of PMTs, and the calibrated gain is used for the photoelectron calculation. The dark count rate (DCR) is also measured under the setting point with a quarter p.e. threshold\footnote{The threshold is also used for the following muon related testing.} here. It is around 16\,kHz for NNVT PMT and 44\,kHz for HPK PMT respectively, where a tiny light leakage in single photon level (around 20-30\,kHz) is found during the testing of HPK PMT from the used black cloth covering of the system.

\begin{figure}[!htb]
	\centering
	\includegraphics[width=0.7\linewidth]{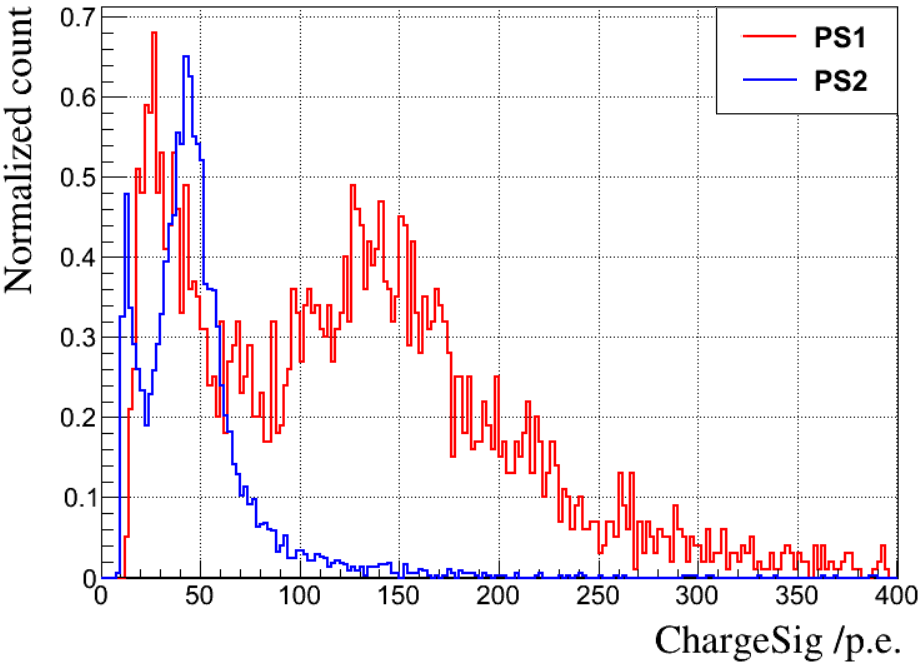}
	\caption{Measured charge spectra of scintillators which are measured with a threshold around 10\,p.e. The response difference comes from their thickness as stated.}
	\label{fig:ps:charge}       
\end{figure}

A similar procedure is applied to the PMTs of the scintillators, a working point is settle down too. The amplitude of SPE is around 4.3\,mV for PS1 and 5.4\,mV for PS2, and the gain is 0.35$\times10^7$ for PS1 and 0.34$\times10^7$ for PS2 respectively. The relationship between the amplitude and charge gain is different to the 20-inch PMTs, which is mainly from the shape features of the pulses as discussed in Sec.\,\ref{2:system}. The count rate of the scintillators with PMT is also measured with an amplitude threshold around 30\,mV (without amplifier), it is around 3\,kHz for PS1 and 1\,kHz for PS2 respectively. The measured charge spectra of the plastic scintillators are shown in Fig.\,\ref{fig:ps:charge}, where the typical effective light yield to muon is around 140\,p.e. for PS1 and 40\,p.e. for PS2. The differences on counting rate and muon response all come from the thickness of the scintillators. Another offline charge cut of scintillaotrs will be used for more purer muon selection: 70\,p.e. for PS1 and 20\,p.e. for PS2\footnote{The threshold is settle done on balance to remove the gammas and to keep a higher muon detection efficiency.}. All the measured parameters are shown in Tab.\,\ref{tab:PMT:setting}.

\begin{table}[!ht]
    \centering
    \caption{Setting point of all the PMTs}
    \label{tab:PMT:setting}       
    \begin{tabular}{l|l|l|l|l}
    \hline\noalign{\smallskip}
    PMT & HV & SPE & Gain & Count rate  \\
        & (V) & amplitude &  ($\times 10^7$)     & (kHz)  \\
        &  & (mV) &  &   \\
    \noalign{\smallskip}\hline\noalign{\smallskip}
    20-inch NNVT & 1770 & 7.0 & 1.20 & 16\\
    20-inch HPK & 1677 & 6.0 & 1.00 & 44\\
    PMT1 w/ PS1  & -2200 & 4.3 & 0.35 & 3\\
    PMT2 w/ PS2  & -1800 & 5.4 & 0.34 & 1\\
    \noalign{\smallskip}\hline
    \end{tabular}
\end{table}

\section{Measurement results}
\label{1:results}

Except the rate of dark count of 20-inch PMT, more features are measured for better understanding on it, such the dark count rate (DCR) versus threshold, the amplitude and charge spectra to understand the sources of dark count, such as sources from thermal emission, muons, natural radioactivity, or flashers. \\
In this section, a detailed measurement on the dark count of 20-inch PMT is done firstly. Then, the dark count related to muons is further measured under several configurations.
All the results will be discussed one by one.

\subsection{Dark count}
\label{2:DN}

\subsubsection{Dark count rate (DCR)}
\label{3:DCR}

\begin{figure}[!htb]
    \centering
    \includegraphics[scale=0.25]{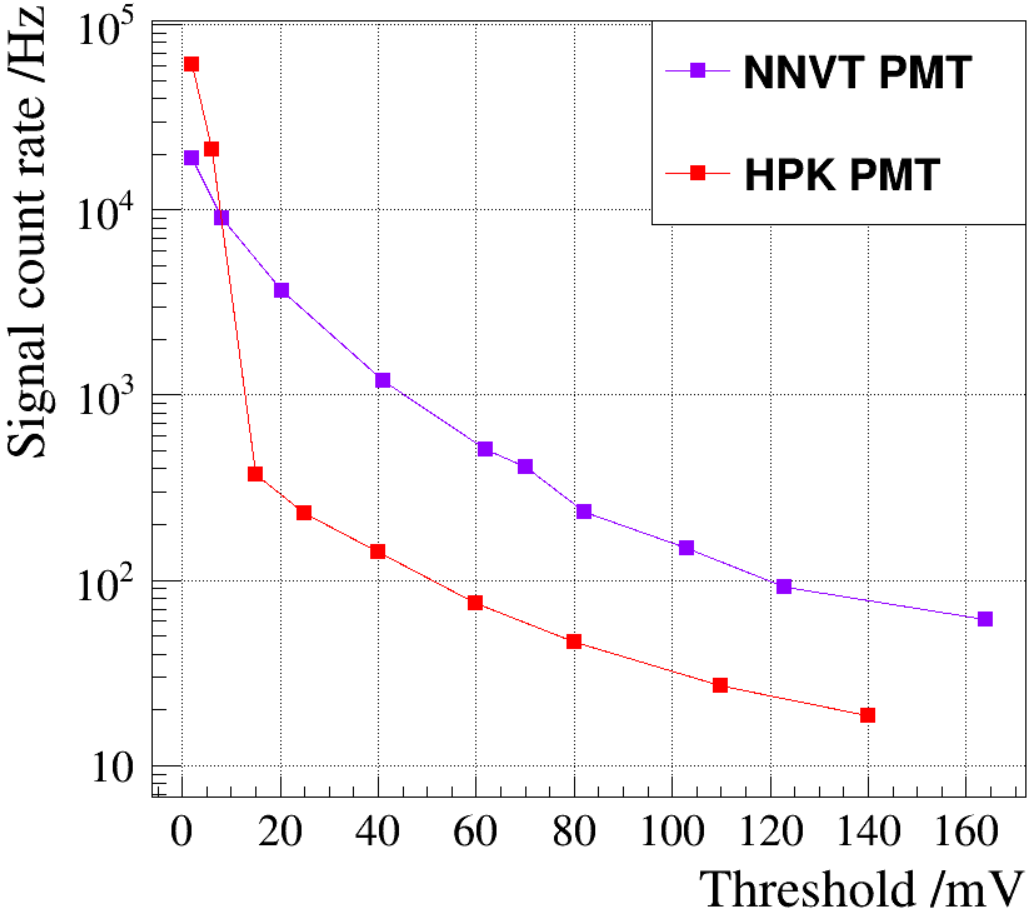}
    \caption{Dark count rate of 20-inch PMTs vs. threshold.}
    \label{fig:PMT:dcr}
\end{figure}

Dark count is mainly source from the thermal electron emission of PMT photocathode in dark. Its amplitude should be in SPE level normally \cite{HamManual,POLYAKOV201369}, which is mostly less than 3\,p.e. and $\ll$1\,Hz for \textgreater 3\,p.e. (assuming DCR in SPE $\sim$ 10\,kHz and 10\,ns coincidence window). A threshold survey is done for the dark count rate of 20-inch PMT as shown in Fig.\,\ref{fig:PMT:dcr}. Due to the tightness of the device, there is a tiny light leakage during HPK PMT measurement as mentioned, which results in a higher count rate when the threshold less than 10\,mV. Comparing with expectation, it is clear that there shows an obvious rate when the threshold is higher than 20\,mV (around 3\,p.e. according to Tab.\,\ref{tab:PMT:setting}), even higher than 50\,mV (around 10 p.e.) in amplitude for both types of PMTs. It is over the contribution only from the thermal noise about the DCR. The higher rate of large pulses of NNVT PMT than HPK PMT is related to the response features of NNVT PMT on amplitude as shown in Fig.\,\ref{fig:PMT:spe}, where the NNVT PMT shows a wider distribution on amplitude than HPK PMT even both of them in SPE level.

\subsubsection{Dark count waveform}
\label{3:DN:wave}

\begin{figure*}[!htb]
	\centering
    \begin{subfigure}[c]{0.495\textwidth}
	\includegraphics[width=\linewidth]{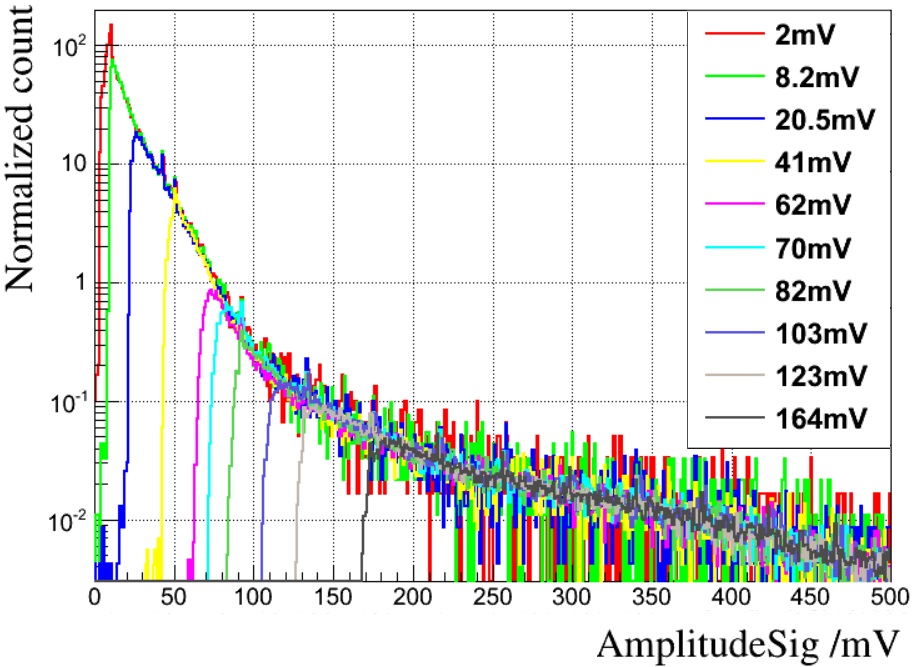}
	\end{subfigure}	\hfill
	\begin{subfigure}[c]{0.495\textwidth}
		\includegraphics[width=\linewidth]{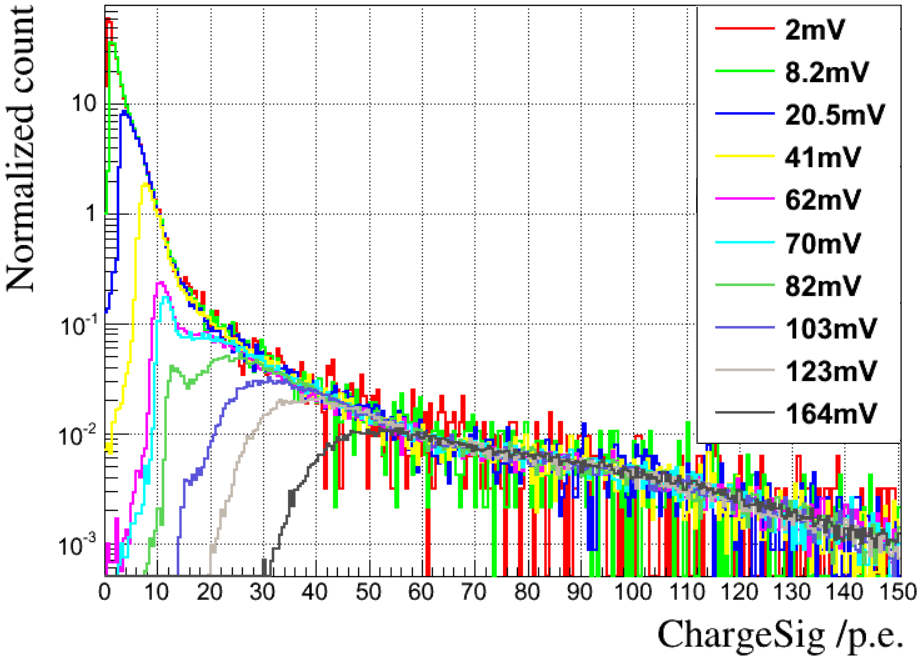}
	\end{subfigure}
	\begin{subfigure}[c]{0.495\textwidth}
	\includegraphics[width=\linewidth]{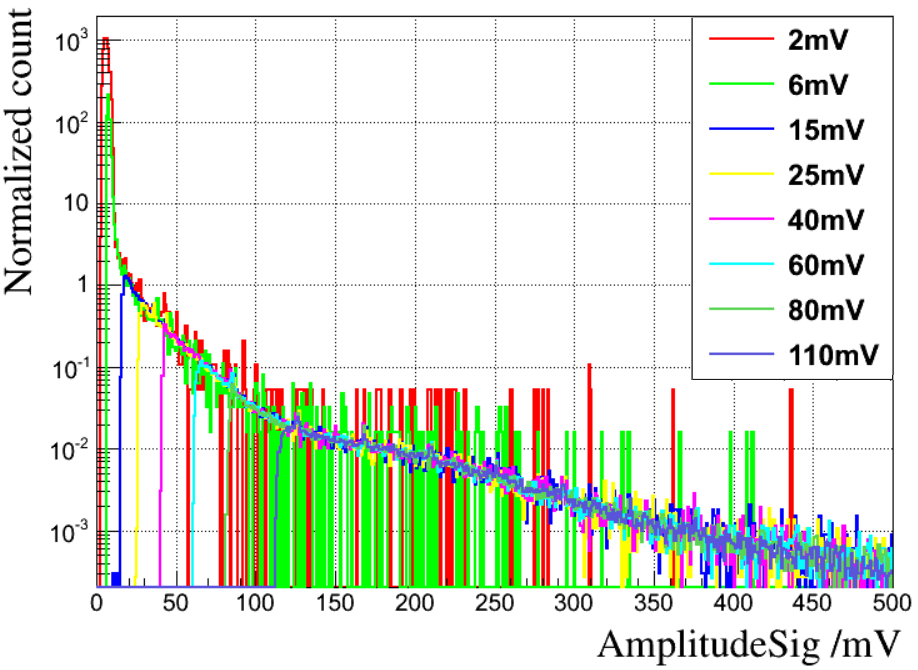}
	\end{subfigure}\hfill
	\begin{subfigure}[c]{0.495\textwidth}
		\includegraphics[width=\linewidth]{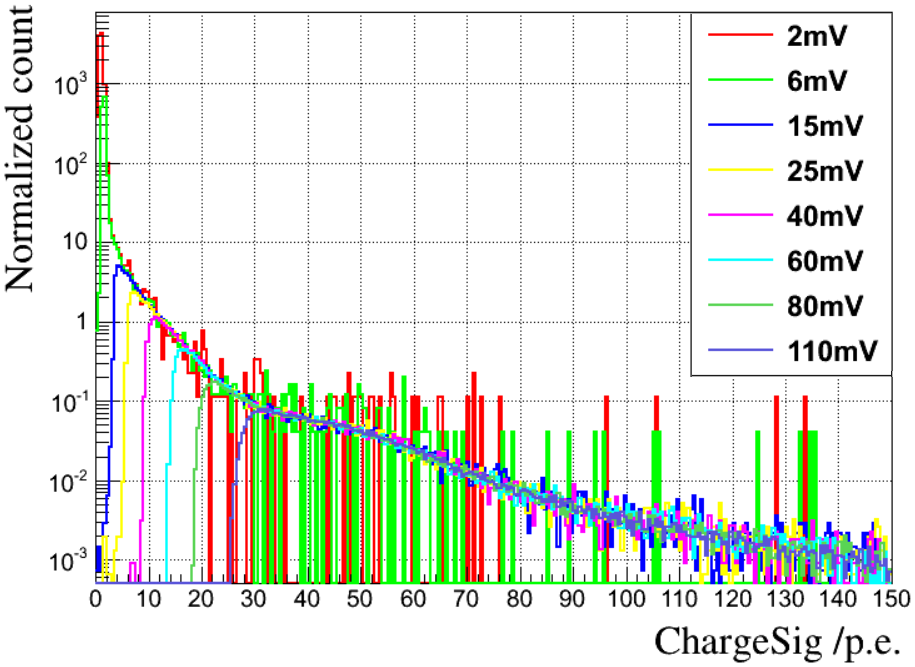}
	\end{subfigure}	
	\caption{The normalized amplitude (left) and charge (right) spectra of 20-inch PMTs’ dark count at different thresholds. Top left: 20-inch NNVT PMT amplitude spectrum; top right: 20-inch NNVT PMT charge spectrum; bottom left: 20-inch HPK PMT amplitude spectrum; bottom right: 20-inch HPK PMT charge spectrum.}
	\label{fig:dn:wave}       
\end{figure*}

The waveforms are further taken during the threshold survey for both types of PMTs with the system\footnote{The data taking dead time is assumed to be random even for high enough trigger rate, and no systematic effect on the overall distribution.}. The distributions of DCR amplitude and charge are shown in Fig.\,\ref{fig:dn:wave}, where all the plots are normalized to the lowest threshold result according to the event numbers higher than 200\,mV. All the spectra of amplitude (charge) of both types of PMTs are following a similar overall trend, and showing a structure in steps, in amplitude (charge):
\begin{itemize}
	\item[1.] The first step from 0 to $\sim$10\,mV (0 to $\sim$3\,p.e. in charge), which should be mainly contributed by the thermal electron emission. The result of HPK PMT suffers from the tiny light leakage.
	\item[2.] The second step from $\sim$10\,mV to $\sim$100\,mV ($\sim$3\,p.e. to $\sim$20\,p.e. in charge), which still needs further understanding.
	\item[3.] The third step from $\sim$100\,mV to $\sim$500\,mV ($\sim$20\,p.e. to at least $\sim$150\,p.e. in charge), which still needs further understanding, which this study is trying to focus on.
\end{itemize}

\subsection{Muon hit PMT glass}
\label{2:muon}

\begin{figure}[!htb]
	\centering
	\begin{subfigure}[c]{0.48\textwidth}
	\centering
	\includegraphics[width=\linewidth]{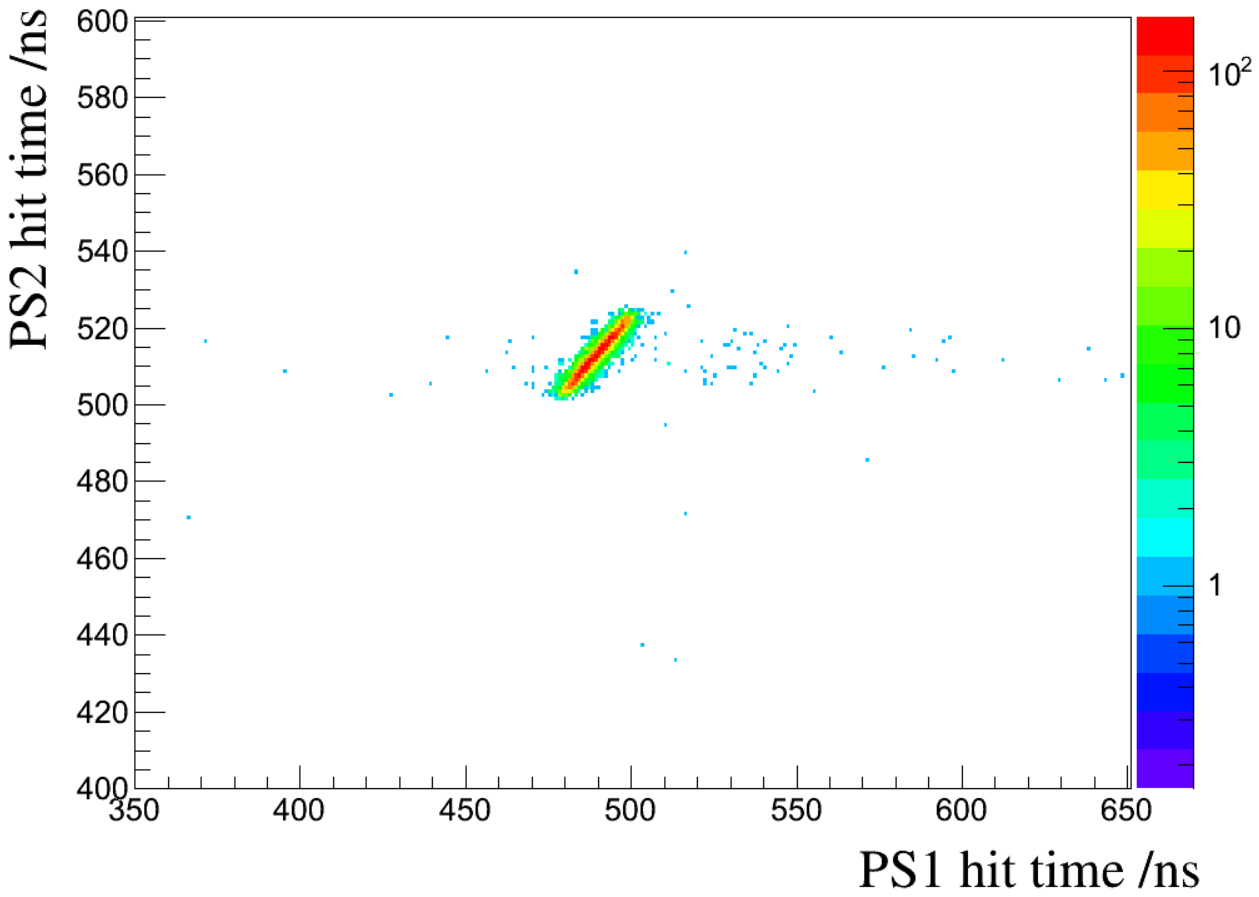}
	\caption{PS1 vs PS2}
	\label{fig:muon:hittime:doublePS}
	\end{subfigure}	
	\begin{subfigure}[c]{0.48\textwidth}
	\centering
	\includegraphics[width=\linewidth]{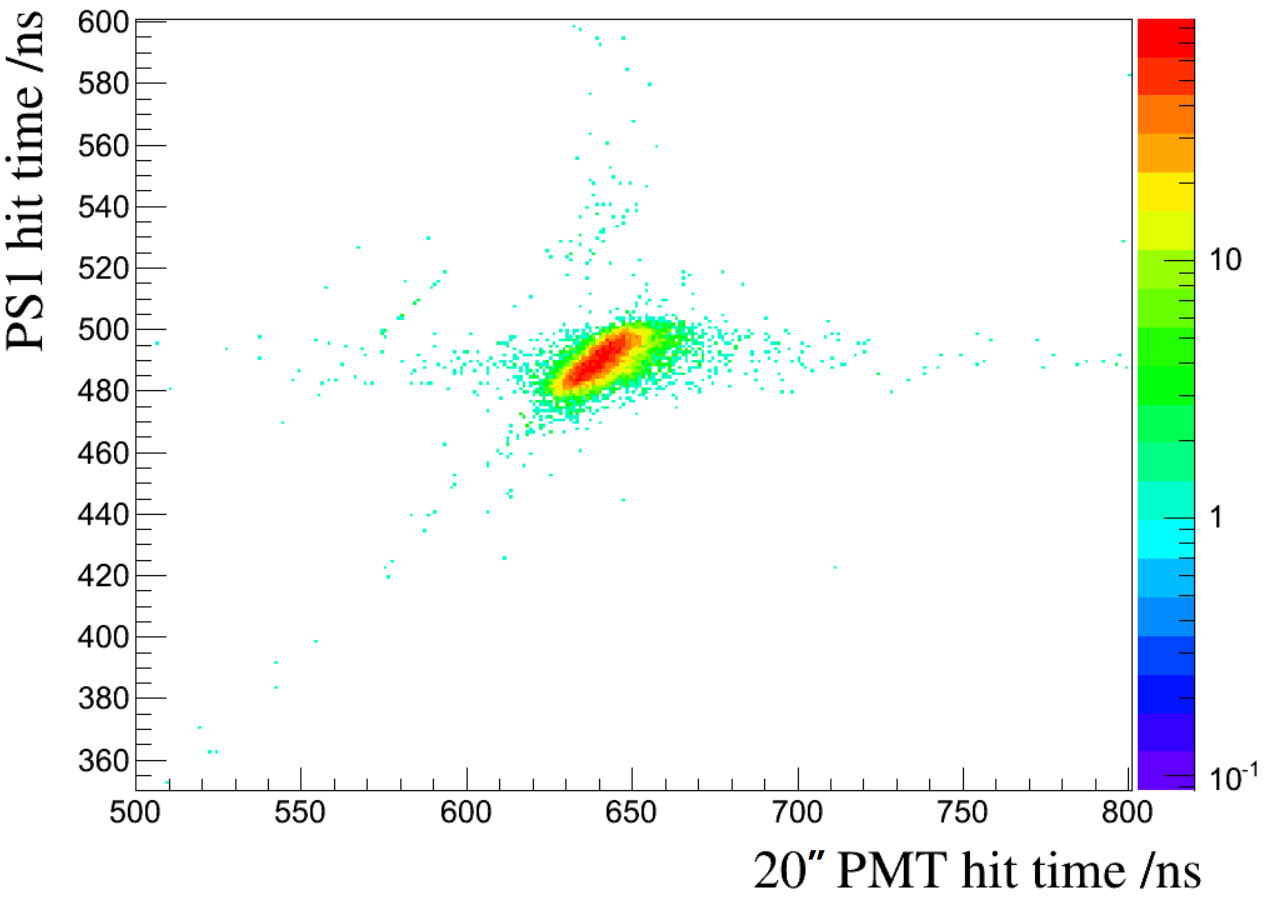}
	\caption{20-inch PMT vs PS1}
	\label{fig:muon:hittime:PMTvsPS}
	\end{subfigure}\hfill
	\caption{2D hit-time correlation between the two scintillators (left) and PS1 and 20-inch PMT (right).}
	\label{fig:hittime}       
\end{figure}

With the system, trigger mode \#3 (triple coincidence of PSs and 20-inch PMT, see Sec.\,\ref{2:system}) is used for muon related signal testing.
The parameters (see Fig.\,\ref{fig:pulse:integration}) are derived from the measured waveforms. The hit-time correlation of both PSs (shown on the left of Fig.\,\ref{fig:hittime}) is checked firstly to exclude possible random noise, where only the events around [470,510]\,ns of x-axis and [500,530]\,ns of y-axis are selected for the following analysis. The hittime correlation between PS1 and 20-inch PMT (shown on the right of Fig.\,\ref{fig:hittime}) is also selected in the similar way for the following analysis.

\begin{figure}[!htb]
	\centering
	\begin{subfigure}[c]{0.45\textwidth}
	\centering
	\includegraphics[width=\linewidth]{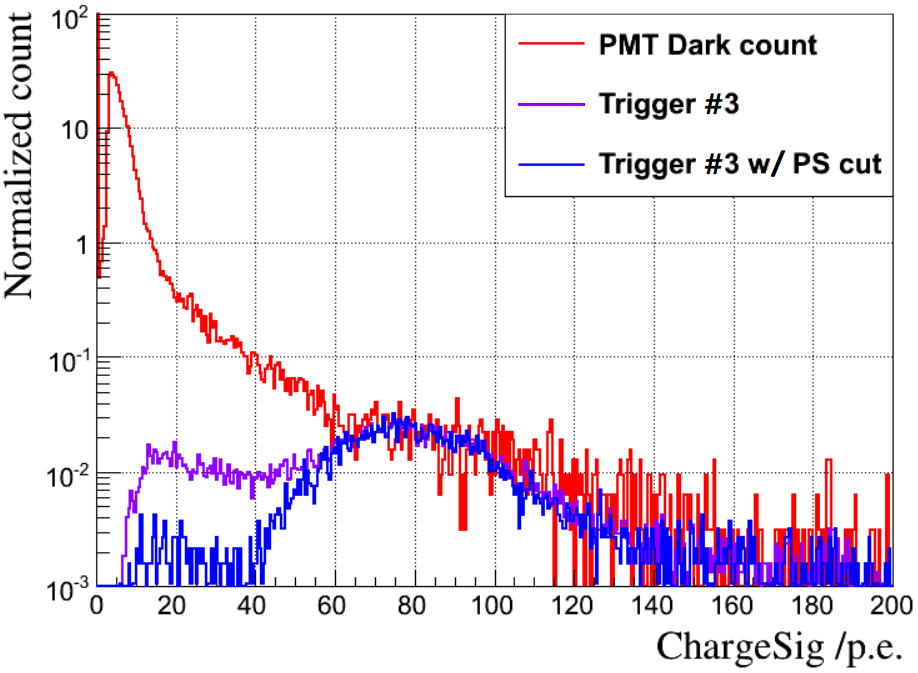}
	\caption{20-inch NNVT PMT}
	\label{fig:mcp:measure:single:NNVT}
	\end{subfigure}	
	\begin{subfigure}[c]{0.45\textwidth}
	\centering
	\includegraphics[width=\linewidth]{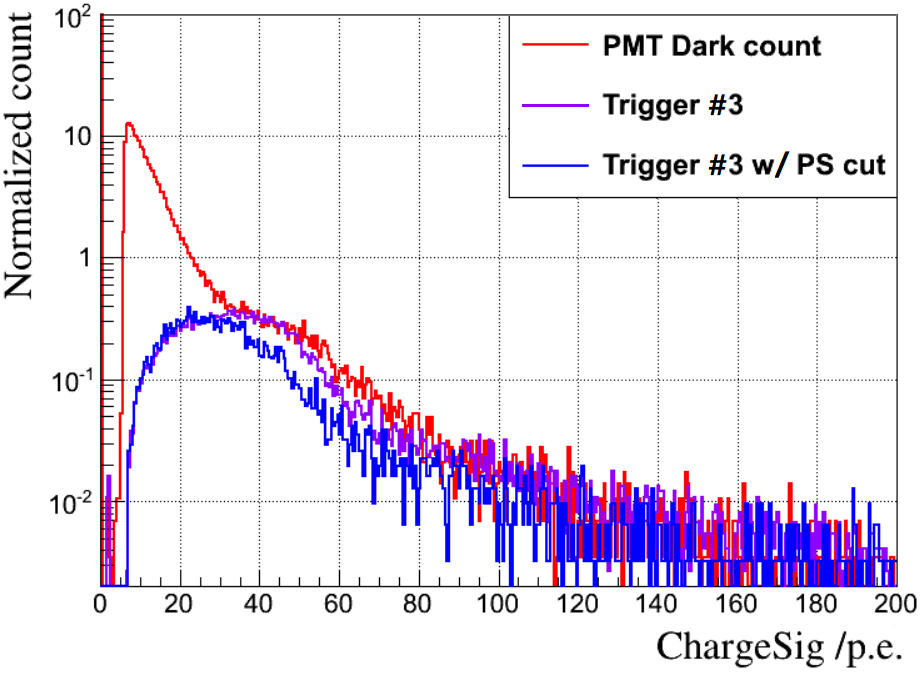}
	\caption{20-inch HPK PMT}
	\label{fig:mcp:measure:single:HPK}
	\end{subfigure}\hfill
	\caption{Measured charge spectra of 20-inch PMTs. Left: MCP PMT; right: HPK PMT. Red: Dark count spectrum (tagged by "PMT dark count"), same as in Fig.\,\ref{fig:dn:wave}; purple: the measured spectrum selected only with the hit time cut of PSs; blue: after offline charge cut.}
	\label{fig:mcp:measure:single}       
\end{figure}

When the PS1 located at the position 0.15\,m ralative to the central axis of 20-inch PMT (zenith angle ($\theta$) of muon incident is around 27$^\circ$ {$\pm$16$^\circ$}), the measured charge spectra after hit-time cut (tagged by "Trigger \#3") of 20-inch MCP PMT can be found on the left of Fig.\,\ref{fig:mcp:measure:single}, where the charge spectrum of dark count (trigger mode \#1, tagged by "PMT dark count") is also shown on the same figure. A further selected spectrum after the offline charge cut of PSs (tagged by "Trigger \#3 w/ PS cut") is drawn on the same figure. All the curves are normalized according to the large pulse events of dark count spectrum. Firstly, it is found on the left of Fig.\,\ref{fig:mcp:measure:single} for NNVT PMT that the muon related events are mainly related to the large pulses with typical charge around 76\,p.e. -- the step 3 as discussed in Sec.\,\ref{3:DN:wave}. Secondly, the muon related events also contribute to part of lower than 40\,p.e. located at the step 2 as discussed in Sec.\,\ref{3:DN:wave}.
Similar features can be identified for the 20-inch HPK PMT as shown on the right of Fig.\,\ref{fig:mcp:measure:single}, where the typical signal intensity in charge is around 30\,p.e. located at the step 3 too.
It is clear that the muon related events mainly contribute to the entries of step 3, and partially contribute to the step 2.

\begin{figure}[!htb]
	\centering
	\begin{subfigure}[c]{0.45\textwidth}
	\centering
	\includegraphics[width=\linewidth]{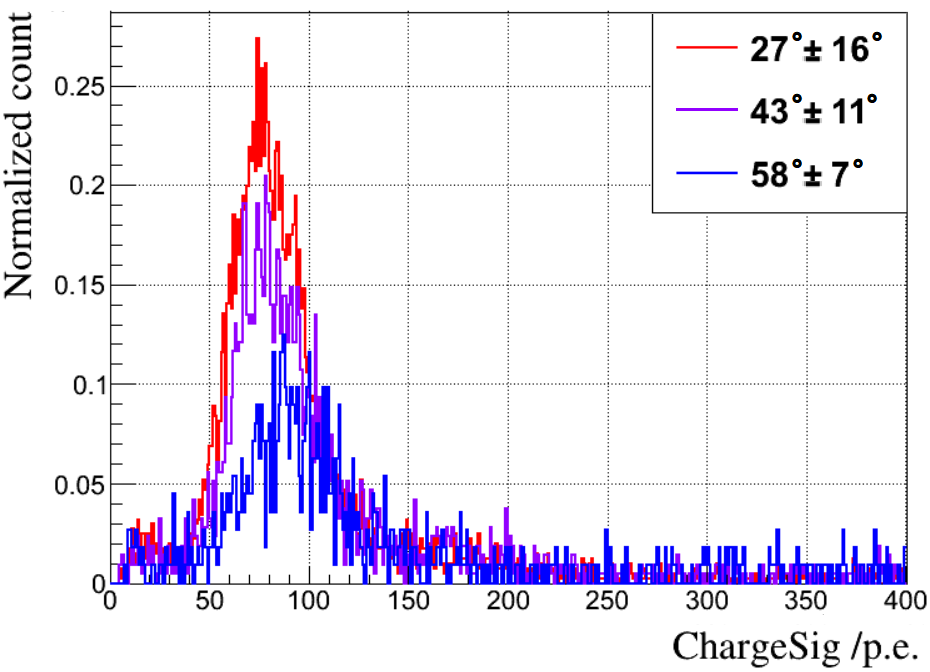}
	\caption{20-inch NNVT PMT}
	\label{fig:mcp:measure:angles:NNVT}
	\end{subfigure}	
	\begin{subfigure}[c]{0.45\textwidth}
	\centering
	\includegraphics[width=\linewidth]{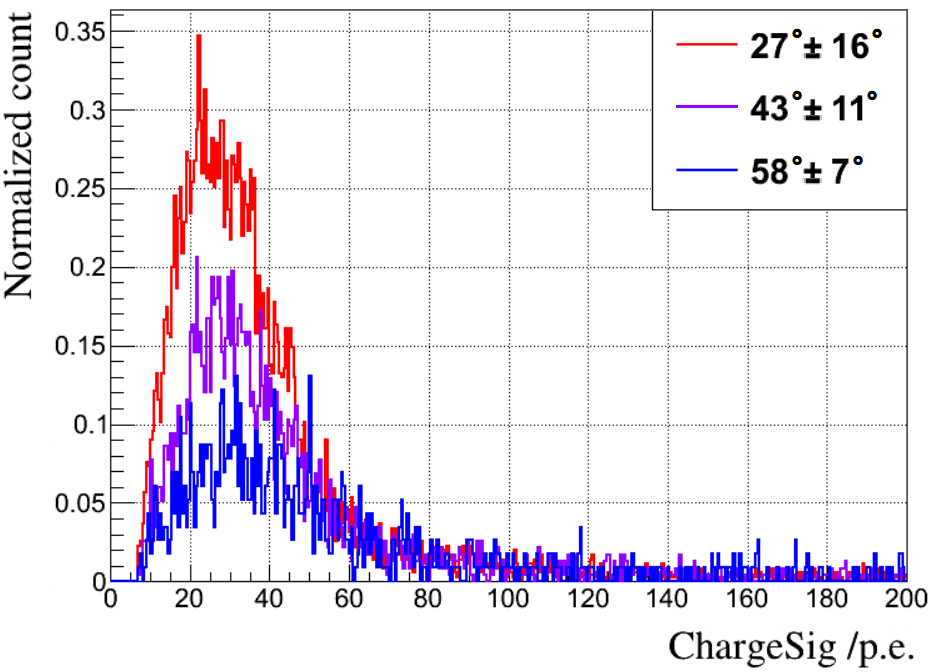}
	\caption{20-inch HPK PMT}
	\label{fig:mcp:measure:angles:HPK}
	\end{subfigure}\hfill
	\caption{Measured charge spectra of 20-inch PMTs in different muon directions. Left: 20-inch NNVT PMT; right: 20-inch HPK PMT. Red: zenith angle ($\theta$) is around 27$^\circ$ $\pm$ 16$^\circ$; purple: zenith angle ($\theta$) is around 43$^\circ$ $\pm$ 11$^\circ$; blue: zenith angle ($\theta$) is around 58$^\circ$ $\pm$ 7$^\circ$. }
	\label{fig:mcp:measure:angles}       
\end{figure}

In order to survey more directions of muons going through the 20-inch PMT, another two data sets are taken when PS1 located at 0.5\,m (zenith angle ($\theta$) is around 43$^\circ$ $\pm$ 11$^\circ$) and 1.0\,m (zenith angle ($\theta$) is around 58$^\circ$ $\pm$ 7$^\circ$) relative to the 20-inch PMT central axis. The measured charge spectra after PSs' offline charge cut are shown in Fig.\,\ref{fig:mcp:measure:angles} for both types of PMTs. All the measurements provide consistent results on the muon related signal strength, while the difference mainly focus on the measured event rate from solid angle. The measured muon rate by PMT only is selected with charge larger than 50\,p.e. for NNVT PMT and larger than 10\,p.e. for HPK PMT, respectively.
The typical charge strength and event rate are collected and summarized in Tab.\,\ref{tab:rate}, where the rate is related to the specified arrangement layout of the location of the two scintillators during each measurement, which could have some uncertainty, especially for PS1.

\begin{table}[!ht]
\centering
\caption{Event strength and rate of different muon directions}
\label{tab:rate}       
\begin{tabular}{l|l|l|l|l}
\hline\noalign{\smallskip}
 Configuration & PMT & rate w/ one PS & rate w/ double PSs & Typical\\
            &          & (w/ PS cut) (Hz) & (w/ PS cut) (Hz) & intensity  \\
            &          & (w/ PS ,w/ PMT cut) & (w/ PS ,w/ PMT cut) & (p.e.) \\
\noalign{\smallskip}\hline\noalign{\smallskip}
\multirow{2}{*}{0.15\,m}  & NNVT & 31.41 (26.27)(12.49) & 1.32 (0.93)(0.89) & 76 \\
\noalign{\smallskip}\cline{2-5}\noalign{\smallskip}
   ($\theta\sim$27$^\circ$ $\pm$ 16$^\circ$) & HPK & 13.12 (9.45)(8.97) & 0.88 (0.59)(0.59) & 30 \\
\noalign{\smallskip}\hline\noalign{\smallskip}
\multirow{2}{*}{0.5\,m }  & NNVT & 16.33 (13.01)(7.66) & 0.65 (0.42)(0.40) & 76  \\
\noalign{\smallskip}\cline{2-5}\noalign{\smallskip}
 ($\theta\sim$43$^\circ$ $\pm$ 11$^\circ$) & HPK & 11.87 (8.06)(7.71) & 0.49 (0.26)(0.26) & 31 \\
\noalign{\smallskip}\hline\noalign{\smallskip}
\multirow{2}{*}{1.0\,m }  & NNVT & 11.25 (8.37)(5.24) & 0.41 (0.26)(0.24) & 85 \\
\noalign{\smallskip}\cline{2-5}\noalign{\smallskip}
 ($\theta\sim$58$^\circ$ $\pm$ 7$^\circ$) & HPK & 8.36 (5.46)(5.12) & 0.34 (0.19)(0.19) & 33 \\
\noalign{\smallskip}\hline
\end{tabular}
\end{table}

\subsubsection{20-inch PMT coincidence only with PS1}
\label{3:PMT:PS1}

\begin{figure}[!htb]
	\centering
	\begin{subfigure}[c]{0.4\textwidth}
	\centering
	\includegraphics[width=\linewidth]{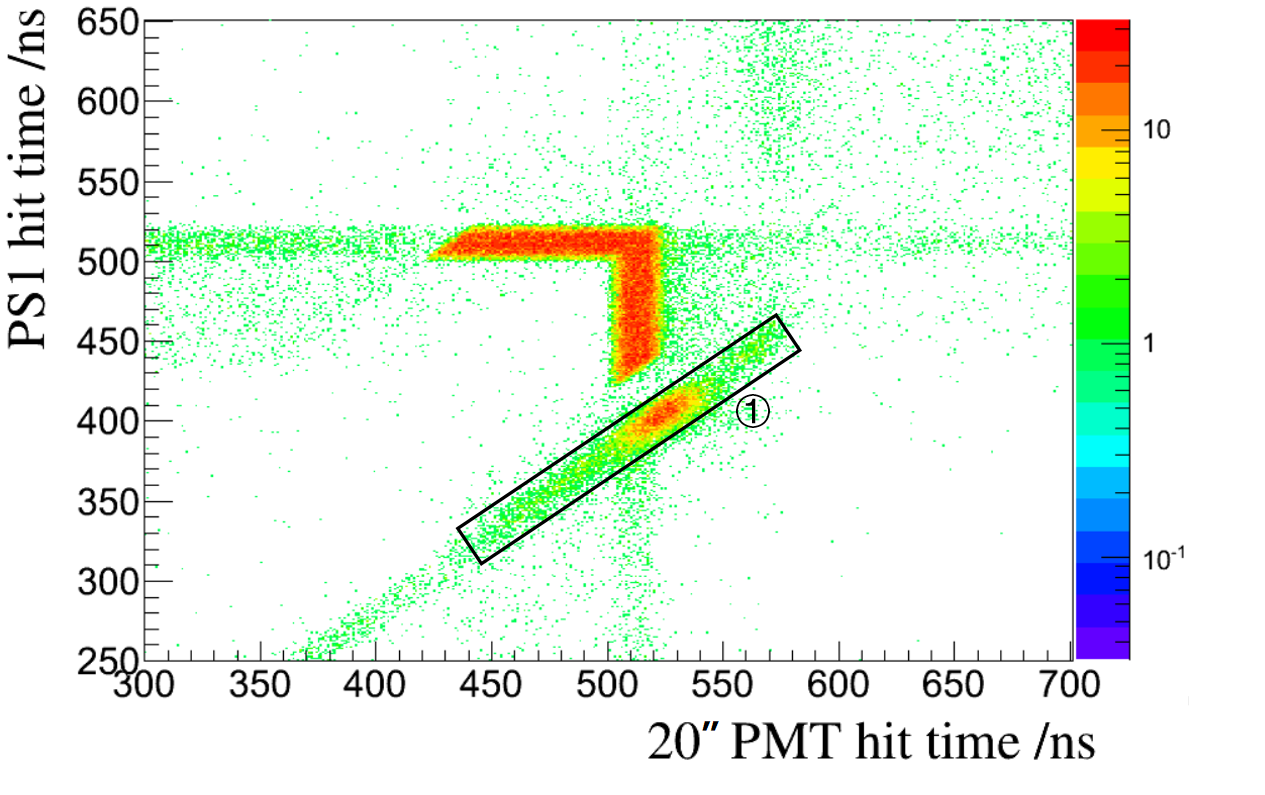}
	\caption{2-D hit-time correlation of 20-inch NNVT PMT and PS1}
	\label{fig:trigger2:2D}
	\end{subfigure}	
	\begin{subfigure}[c]{0.35\textwidth}
	\centering
	\includegraphics[width=\linewidth]{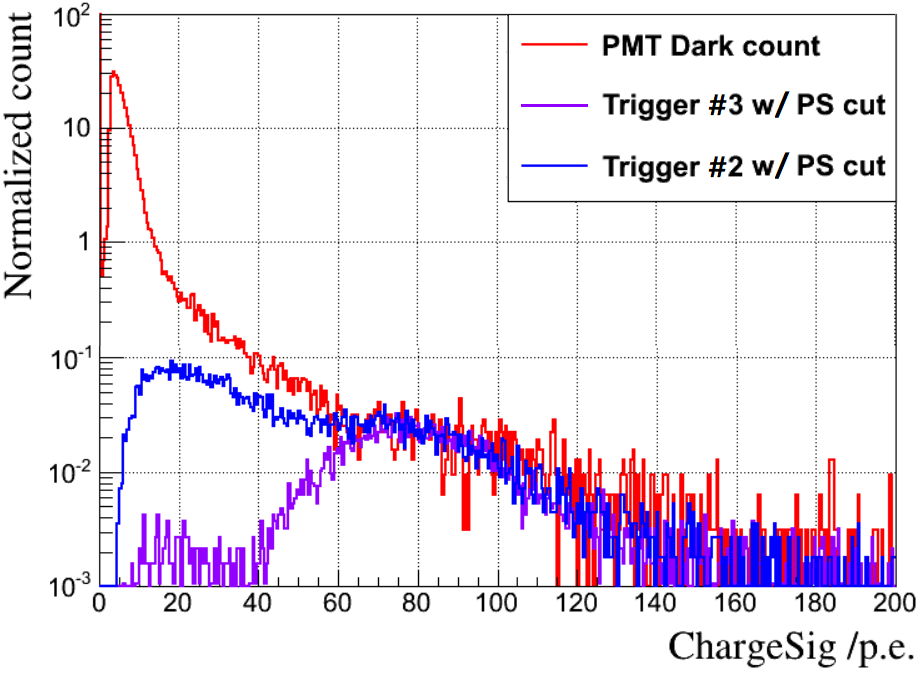}
	\caption{Measured charge spectra of 20-inch NNVT PMT.}
	\label{fig:trigger2:charge}
	\end{subfigure}\hfill
	\caption{2-D hit-time correaltion between 20-inch NNVT PMT and the PS1 of trigger mode \#2 (left) and the measured charge spectra of 20-inch PMTs (right).}
	\label{fig:1ps}       
\end{figure}

Another configuration with trigger mode \#2 (20-inch PMT coincidence only with PS1, see Sec.\,\ref{2:config}) is used for the measurements with 20-inch PMT for cross check of muon related events, where PS1 is located at 0.15\,m relative to the PMT central axis. The 2-D hit-time correlation between the 20-inch PMT and PS1 is shown on the left of Fig.\,\ref{fig:1ps}, which shows about three regions\footnote{The entries out of the aimed hit-time window are mainly from the random coincidence and secondary pulses.} and a more complex pattern than that of trigger mode \#3 as shown in Fig.\,\ref{fig:hittime}. The regions 2 and 3 on the 2-D hit-time plot are from the random coincidence between the 20-inch PMT and the PS1, and region 1 is the target range generated by the events going through the 20-inch PMT and the PS1 at the same time, which is confirmed by the relative hit-time difference distribution. The charge spectrum measured with trigger mode \#2 of hit-time region 1 is shown on the right of Fig.\,\ref{fig:1ps}, where all the curves are normalized to the spectrum of PMT dark count. All the plots show consistent results for the muon related events, while the lower part of trigger mode \#2 has an obvious increase. It is possibly relate to the natural radioactivity or muon hitting on the PMT glass. Based on the results, it is possible that the step 2 is mainly contributed by the muon and natural radioactivity of the surroundings.

\subsubsection{PMT photocathode down}
\label{3:PMTdown}

\begin{figure}[!htb]
	\centering
	\includegraphics[width=0.6\linewidth]{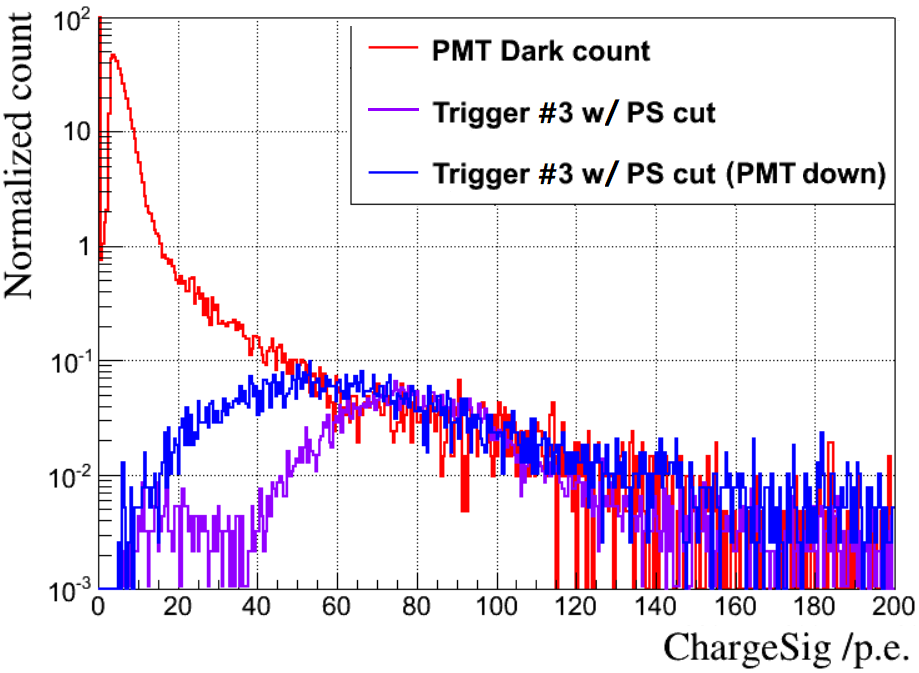}
    \caption{Measured charge spectra of 20-inch NNVT PMT when its photo-cathode towards to down.}
	\label{fig:pmt:down}       
\end{figure}

Concerning the correlation between the Cerenkov light direction and photocathode acceptance when a muon going through the glass bulb, another special test is done further with the PMT photocathode towards to down (comparing towards to up). The trigger mode \#3 is also used for the test only with 20-inch NNVT PMT. The measured charge spectra are shown in Fig.\,\ref{fig:pmt:down} with the raw spectra and after further PSs' offline cut. It also shows obvious muon related events, which is basically consistent with the photocathode directed upwards. But the typical intensity of photocathode towards to down is around 50\,p.e. which is only about 66\% of photocathode towards to up. The direction of Cerenkov light will affect the PMT acceptance for a muon passing through the PMT glass bulb.

\section{Simulation}
\label{1:sim}

 Geant4\cite{geant4} is widely used for particle physics detector simulation. A Geant4 simulation of the whole setup with muons passing through the PMT glass bulb was done and compared with the experimental measurements. The results will help on the understanding of the measurements and to prospect the features of this process in the future JUNO detector.

\subsection{Simulation setup}
\label{2:geo}

\begin{figure}[!htb]
	\centering
	\begin{subfigure}[c]{0.4\textwidth}
	\centering
	\includegraphics[width=\linewidth]{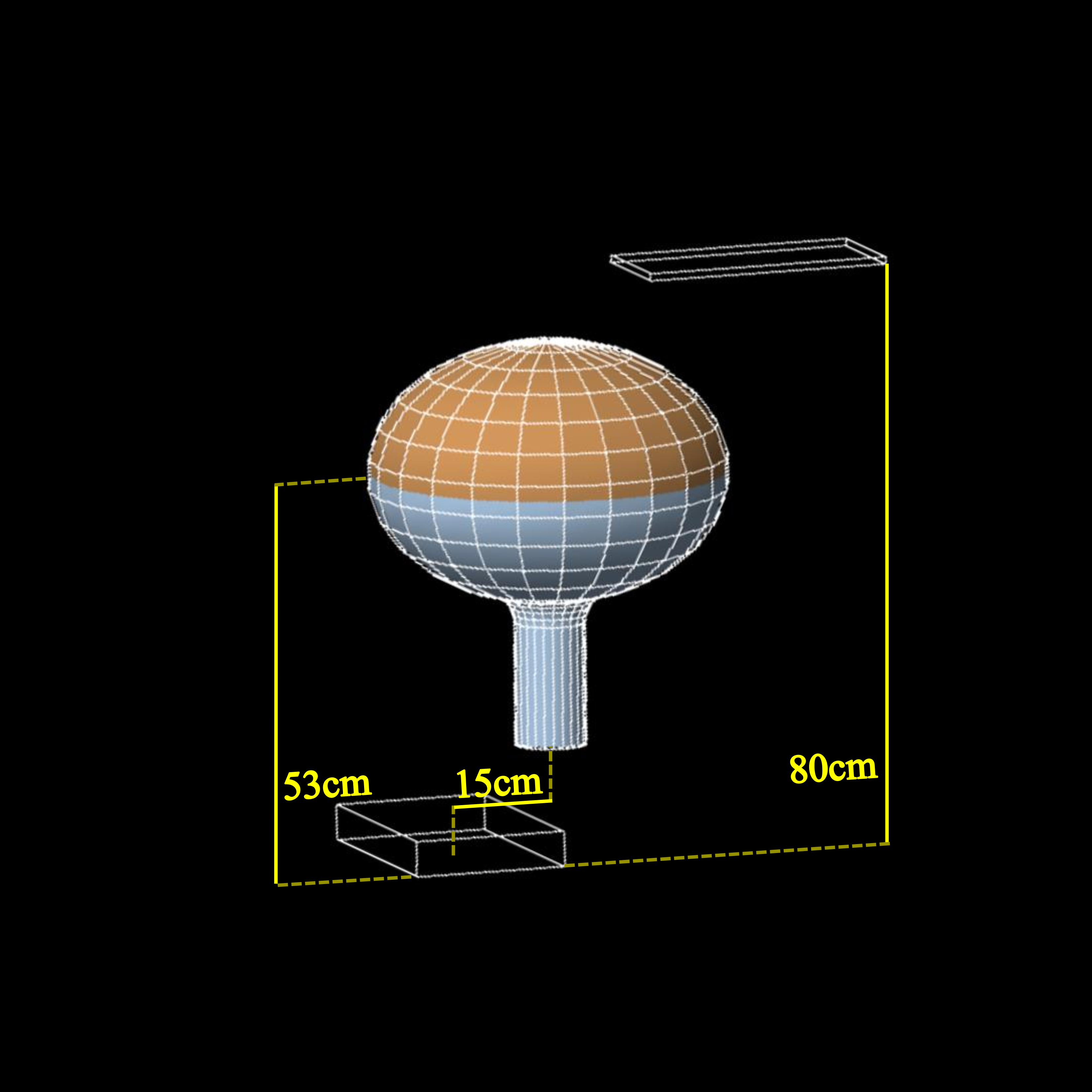}
	\caption{NNVT}
	\label{fig:sim:geo:NNVT}
	\end{subfigure}	
	\begin{subfigure}[c]{0.4\textwidth}
	\centering
	\includegraphics[width=\linewidth]{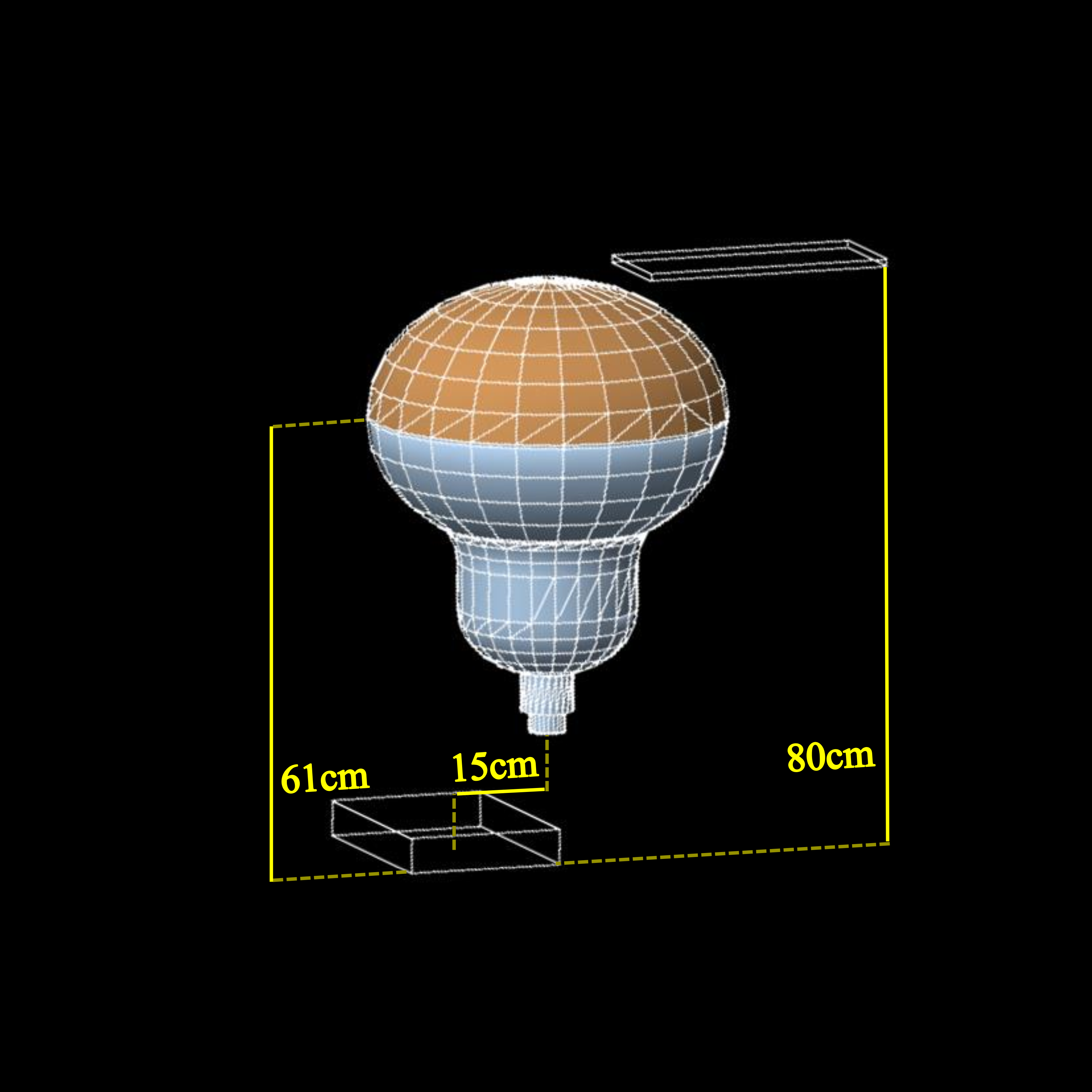}
	\caption{HPK}
	\label{fig:sim:geo:HPK}
	\end{subfigure}\hfill
	\caption{Geometry realized in the simulation for 20-inch NNVT PMT (left), HPK PMT (right) and two plastic scintillators based on Geant4.}
	\label{fig:sim:geo}       
\end{figure}

The simulation project is setup with the 20-inch PMT (HPK or NNVT) and two plastic scintillators inside a standalone dark box, where all the parameters are configured following the study in \cite{Lin_2017}. The realized geometry can be found in Fig.\,\ref{fig:sim:geo}. The muon generator is based on the muon momentum distribution at sea level \cite{Muonflux-Guan2015APO}.

\begin{figure}[!htb]
	\centering
	\begin{subfigure}[c]{0.41\textwidth}
	\centering
	\includegraphics[width=\linewidth]{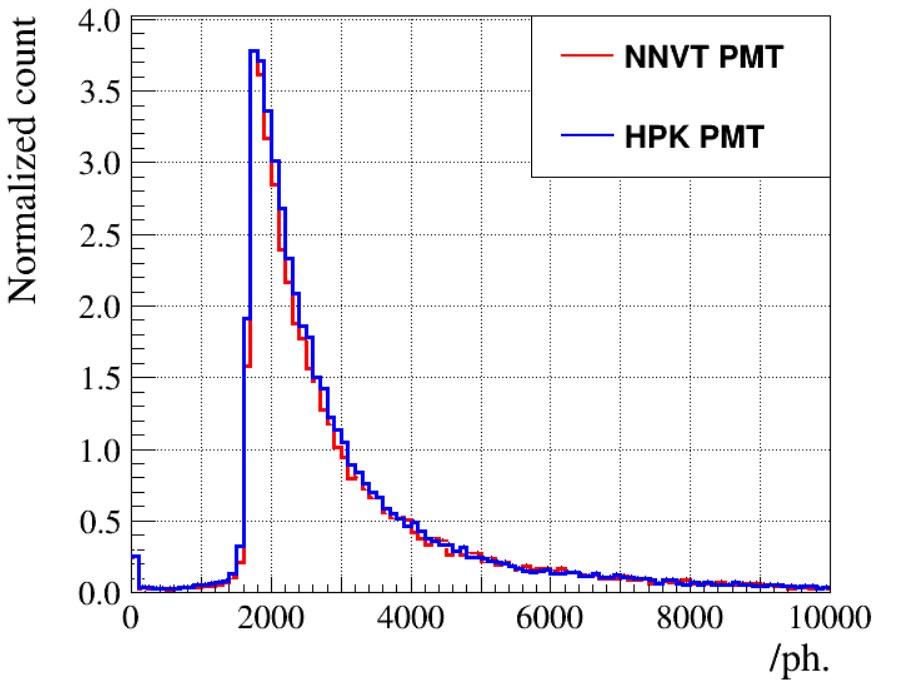}
	\caption{Generated photon by muons corresponds to 6\,mm glass (including in and out).}
	\label{fig:sim:gen:ph:num}
	\end{subfigure}
	\begin{subfigure}[c]{0.41\textwidth}
	\centering
	\includegraphics[width=\linewidth]{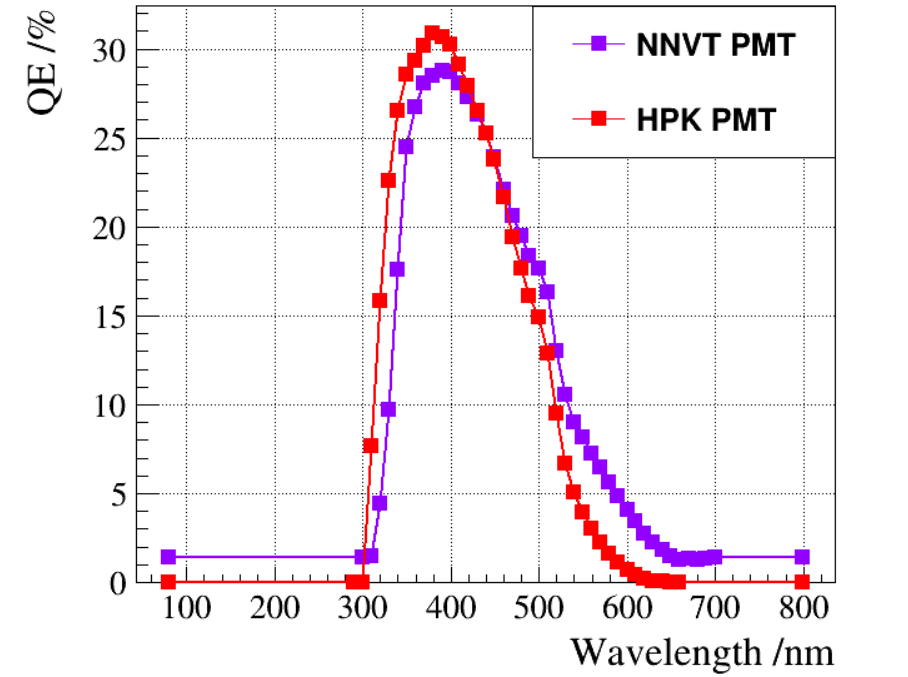}
	\caption{QE curves for both NNVT and HPK 20-inch PMTs}
	\label{fig:sim:gen:ph:QE}
	\end{subfigure}
	\caption{Generated photon number by muons (left) and input QE curves (right).}
	\label{fig:sim:input}       
\end{figure}

There are few critical configurations to mention:
\begin{itemize}
	\item[(1)] Construct the geometry models of PS1, PS2, 20-inch NNVT PMT and HPK PMT, and realize optical processes for Cerenkov radiation in PMT glass. An uniform thickness of PMT glass is used in the simulation.
	\item[(2)] Verify the simulation process related to the PMT glass including Cerenkov radiation (see Fig.\,\ref{fig:sim:gen:ph:num}) and QE curve of photocathode (see Fig.\,\ref{fig:sim:gen:ph:QE}).
	\item[(3)] Muons are generated following the distribution with 4$\pi$ solid angle randomly on a plane 10$\times$10\,m$^{2}$ just above PS2.
\end{itemize}

Photons will be generated by Cerenkov radiation when a muon is passing through the PMT glass if its speed exceeds the phase velocity of light in the glass. The produced photon number depends on the refractive index (n) of the passing through medium, here the refractive index model of glass. Assume the glass index n=1.5 in the wavelength range, the distribution of generated photon number by muon within the wavelength range of 80\,nm-800\,nm is shown in Fig.\,\ref{fig:sim:gen:ph:num}, where the photon production yield is about 274\,ph./mm (about 26\,ph./mm within 400\,nm-700\,nm), basically consistent with theoretical calculation.

The used QE curves of the 20-inch PMT\cite{PMTrelativeCE,largePMTlei} are shown in Fig.\,\ref{fig:sim:gen:ph:QE}, where the value of QE at 420\,nm is used as a normalization factor (the shape of the QE curve is unchanged.) to check for relative effect of QE. The QE of NNVT PMT shorter than 300\,nm is higher than that of HPK PMT, which will result in a higher output strength of NNVT than HPK PMT. In default, the uniform glass thickness is set to 3\,mm for both types of PMTs, and the QE at 420\,nm is set to 27.3\% for NNVT PMT and 30.8\% for HPK PMT respectively.

The typical generated number of photons is around 1771-1793 (100\%) in wavelength range of [80,800]\,nm (considering muon in and out of the close shape of the bulb) when a muon going through the PMT glass. The photon number of hitting the photocathode is around 1246-1267 (70\%), which is contributed a lot by total reflection from glass to air (outside PMT volume). The final strength on photocathode of NNVT after QE conversion is 45-50\,p.e. (2.8\%) and that of HPK is 30-36\,p.e. (2.0\%). As mentioned, the difference between the two types is mainly source from the shape of QE curve, especially the wavelength range of 80-320\,nm.

\subsubsection{Effect of glass thickness and QE}
\label{3:factorcheck}

\begin{figure}[!htb]
	\centering
	\begin{subfigure}[c]{0.45\textwidth}
	\centering
	\includegraphics[width=\linewidth]{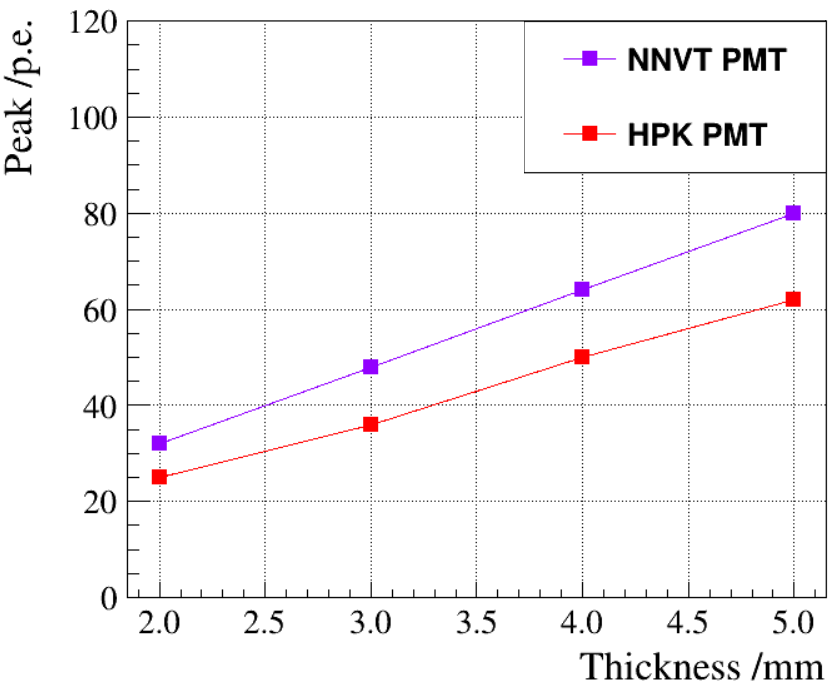}
	\caption{Effect of glass thickness}
	\label{fig:sim:thickness}
	\end{subfigure}	
	\begin{subfigure}[c]{0.45\textwidth}
	\centering
	\includegraphics[width=\linewidth]{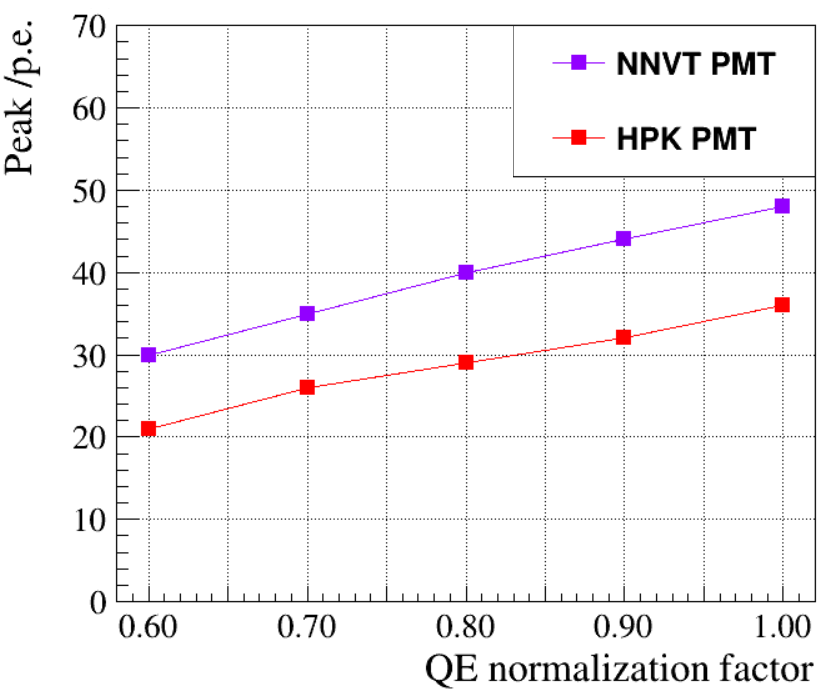}
	\caption{Effect of QE effect}
	\label{fig:sim:QE}
	\end{subfigure}\hfill
	\caption{Signal strength from muon versus glass thickness (left) and QE (right) of 20-inch PMT from simulation. The x-axis of right plot is the ratio to default QE.}
	\label{fig:sim:factorcheck}       
\end{figure}

The glass thickness and QE coefficient of 20-inch PMT have a significant effect on the detected signal strength from the muon related Cerenkov radiation. A survey on glass thickness and QE normalization factor is done individually, as shown in Fig.\,\ref{fig:sim:factorcheck}. The output strength of muon related signal is nearly linear to the glass thickness and the QE: $\sim$16\,p.e./mm for NNVT PMT and $\sim$13\,p.e./mm for HPK PMT respectively; $\sim$8\,p.e./2.7\% for NNVT PMT and $\sim$6\,p.e./2.7\% for HPK PMT, the difference is mainly from the difference of QE shape of the two PMT types.

\subsubsection{Simulation for muons}
\label{3:muonsimulation}

\begin{figure}[!htb]
    \centering
	\begin{subfigure}[c]{0.45\textwidth}
	\centering
	\includegraphics[width=\linewidth]{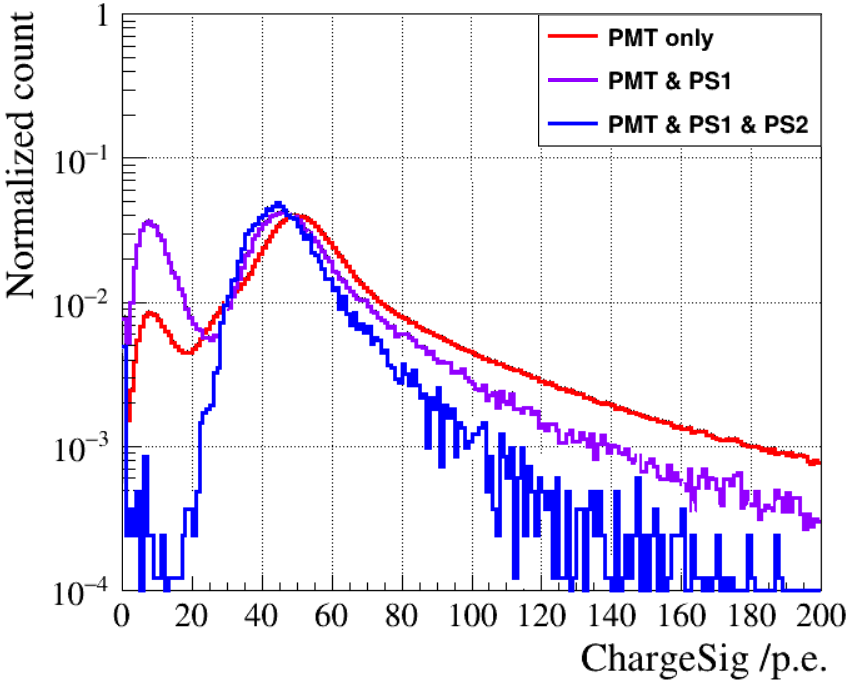}
	\caption{Trigger modes of NNVT PMT}
	\label{fig:sim:nnvt:select}
	\end{subfigure}	
	\begin{subfigure}[c]{0.45\textwidth}
	\centering
	\includegraphics[width=\linewidth]{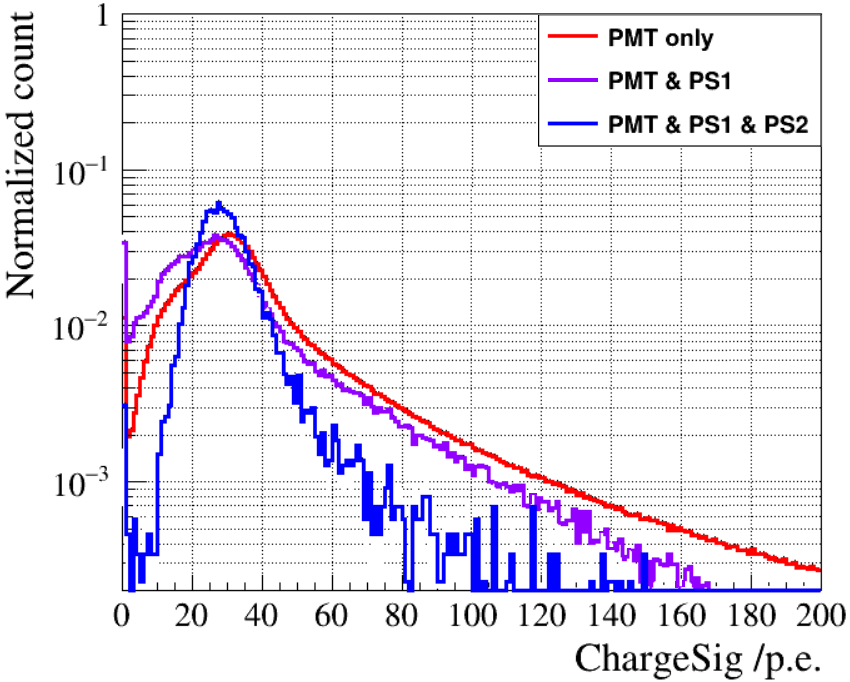}
	\caption{Trigger modes of HPK PMT}
	\label{fig:sim:hpk:select}
	\end{subfigure}
	\begin{subfigure}[c]{0.45\textwidth}
	\centering
	\includegraphics[width=\linewidth]{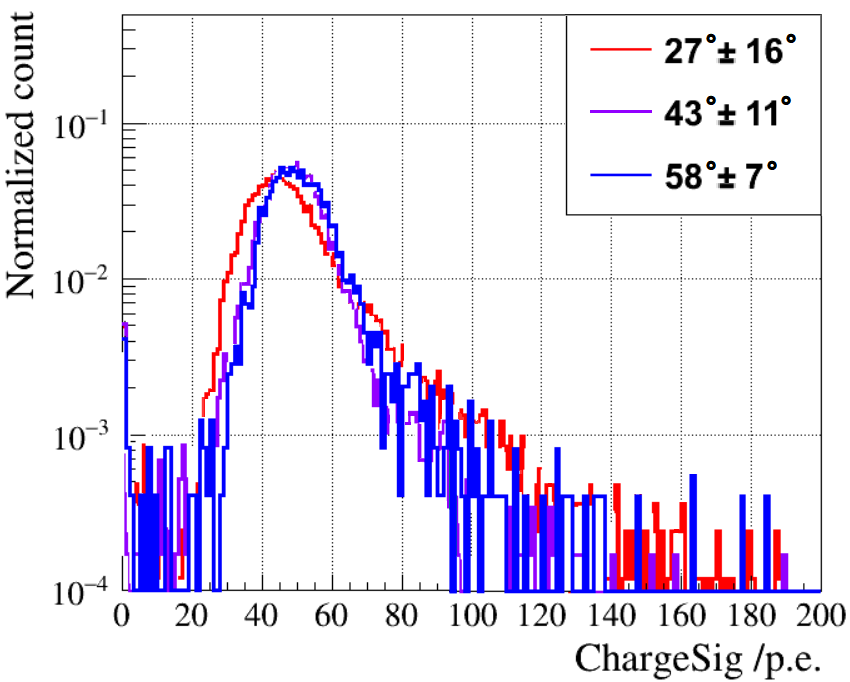}
	\caption{Muon directions of NNVT PMT.}
	\label{fig:sim:nnvt:all}
	\end{subfigure}	
	\begin{subfigure}[c]{0.45\textwidth}
	\centering
	\includegraphics[width=\linewidth]{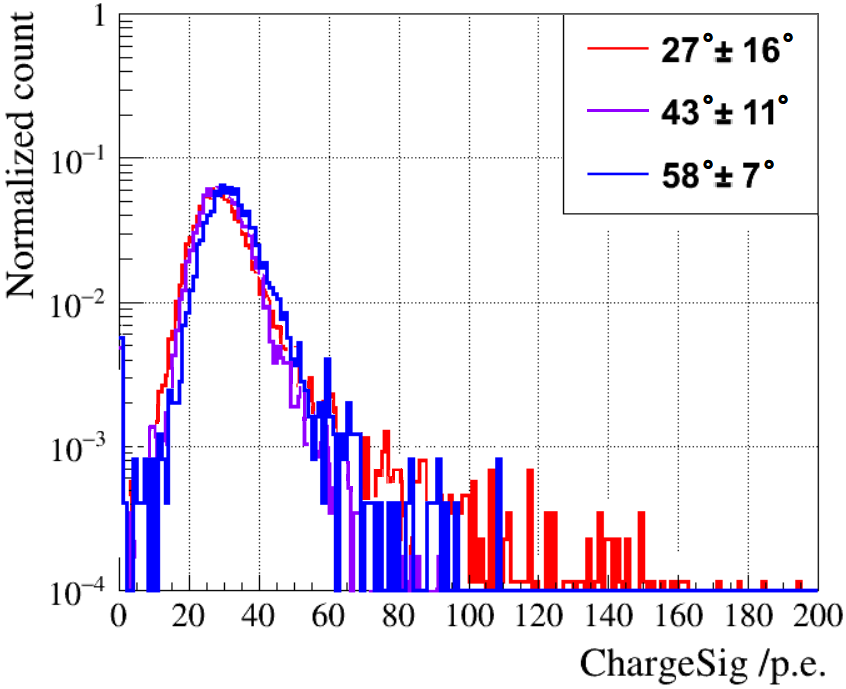}
	\caption{Muon directions of HPK PMT}
	\label{fig:sim:hpk:all}
	\end{subfigure}
	\caption{Typical output from simulation of 20-inch PMTs. Top left: simulation of NNVT PMT for 27$^{\circ}\pm$16$^{\circ}$; top right: simulation of HPK PMT for 27$^{\circ}\pm$16$^{\circ}$; bottom left: simulated signal strength of different angles of NNVT PMT; bottom right: simulated signal strength of different angles of HPK PMT.}
	\label{fig:sim:spec}       
\end{figure}

The simulation output is consistent among different trigger modes for both types of PMTs as shown in Fig.\,\ref{fig:sim:nnvt:select} for NNVT PMT and Fig.\,\ref{fig:sim:hpk:select} for HPK PMT with 27$^{\circ}\pm$16$^{\circ}$, where all the curves are scaled to a similar statistics around the typical peak. The ratio to only the 20-inch PMT (trigger mode \#1) is around 1.9-2.9\% with one PS (trigger mode \#2) and 0.2\% with two PSs (trigger mode \#3) respectively. The simulated distribution of detected signal strength by photocathode for different muon angles can be found in Fig.\,\ref{fig:sim:nnvt:all} for NNVT PMT and Fig.\,\ref{fig:sim:hpk:all} for HPK PMT with an artificial scale too. The typical values of simulation and measurements are list in Tab.\,\ref{tab:comparison}, where the original muon rate in simulation is assumed to 200\,Hz/m$^2$ at ground.

\begin{figure}[!htb]
	\centering
	\begin{subfigure}[c]{0.45\textwidth}
	\centering
	\includegraphics[width=\linewidth]{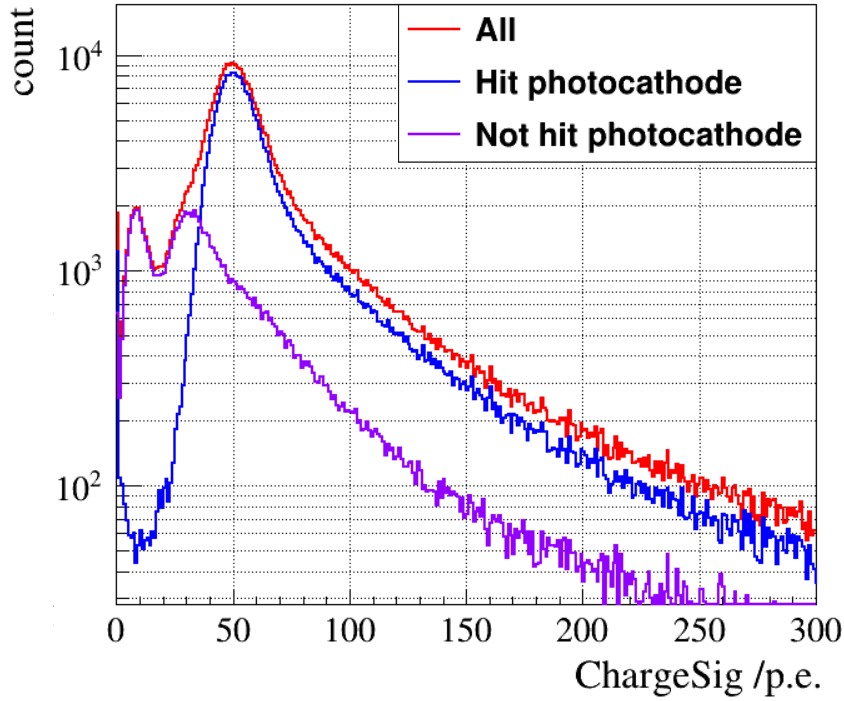}
	\caption{20-inch NNVT PMT.}
	\label{fig:sim:hitlocation:nnvt}
	\end{subfigure}	
	\begin{subfigure}[c]{0.45\textwidth}
	\centering
	\includegraphics[width=\linewidth]{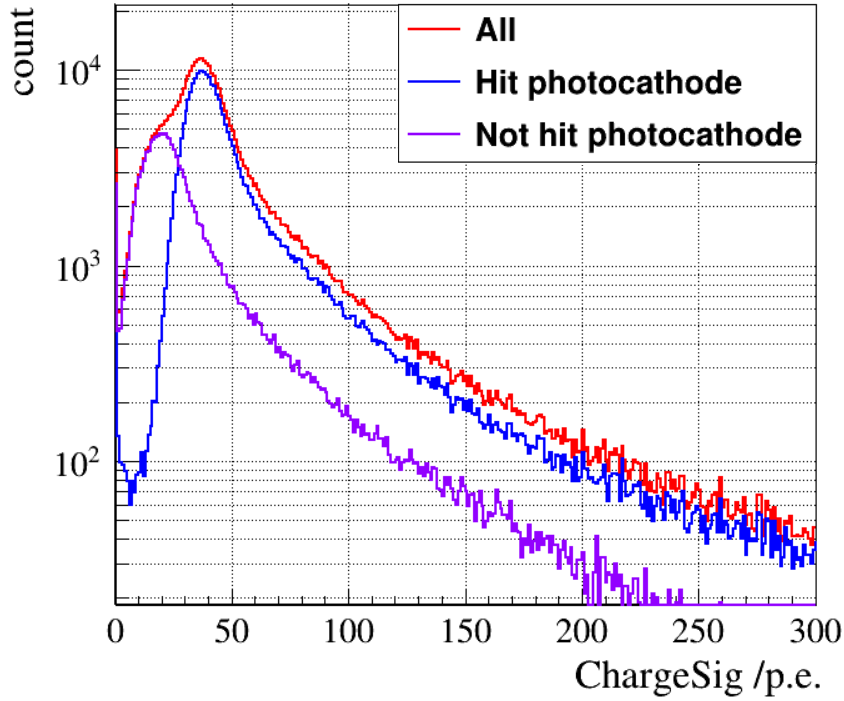}
	\caption{20-inch HPK PMT.}
	\label{fig:sim:hitlocation:hpk}
	\end{subfigure}
	\caption{Effect of muon hitting location: passing through or not the photocathode of 20-inch PMTs.}
	\label{fig:sim:hitlocation}       
\end{figure}

Fig.\ref{fig:sim:hitlocation} is trying to identify the effect of muon direction or muon hitting location on PMT glass shown in Fig.\,\ref{fig:sim:nnvt:select} and Fig.\,\ref{fig:sim:hpk:select} further. The muons hitting the PMT glass are separated into two groups: the muons hitting the PMT photocathode (tagged by "Hit photocathode") and the muons not hitting the PMT photocathode (tagged by "Not hit photocathode"). It shows that even the muon is not hitting the PMT glass with photocathode directly (only passing though the other end of the glass bulb), the PMT still can "see" the photons from the muon. The total refection effect of PMT glass is helping in this case. The shape difference of the charge spectra between NNVT and HPK PMTs for "Not hit photocathode" is mainly source from the different shape of PMT glass bulb of the two types of PMTs as shown in Fig.\ref{fig:sim:geo}.

\begin{table}[!htb]
\centering
\caption{Comparison of simulation and measurement at zenith angle $\theta = $27$^\circ$ $\pm$ 16$^\circ$ of trigger mode \#3. The rate of the simulation is counted in  three columns: the total muon hit rate (tagged by "all"), the muon hit rate passing the photocathode (tagged by "photo-") and the muon hit rate not passing the photocathode (tagged by "not photo-"). The rate of PMT only of the measurement is counted only in the range of >40\,p.e. for NNVT, >10\,p.e. for HPK, with the PS offline cut.}
\label{tab:comparison}       
\resizebox{\linewidth}{!}{
\begin{tabular}{l|l|l|l|l|l|l|l}
\hline\noalign{\smallskip}
 PMT & Config. & Sim. & Test & Sim. rate & Sim. rate  & Sim. rate  & Test rate\\
            & (trig. mod.)  & (p.e.) & (Cor.) (p.e.) & (all)(Hz) & (photo-)  & (not photo-)  & (Hz) \\
\noalign{\smallskip}\hline\noalign{\smallskip}
\multirow{6}{*}{NNVT}  & 20" only & 50 & / (/) & 77.98 &  56.86  & 21.12 & 69.33\\
     &  (\#1)&   &   &   &  &   &  \\
\noalign{\smallskip}\cline{2-8}\noalign{\smallskip}
     & 20" \& PS1  & 47 & 79  & 2.24 & 1.41 & 0.83 & 12.49 \\
    & (\#2) & & (49) &  & &   &  \\
\noalign{\smallskip}\cline{2-8}\noalign{\smallskip}
     & 20" \& PS1 & 45 & 76  & 0.18 & 0.17 & 0.01 & 0.89\\
    &  \& PS2 (\#3) &  & (47.5) &  & &   & \\
\noalign{\smallskip}\hline\noalign{\smallskip}
\multirow{6}{*}{HPK} & 20" only  & 36 & / & 87.04 & 57.57 & 29.47 & 75.19\\
 &  (\#1)&   &   &   & &   &  \\
\noalign{\smallskip}\cline{2-8}\noalign{\smallskip}
     & 20" \& PS1 & 32 & 34 & 1.69 & 1.03  & 0.66 & 8.97\\
     & (\#2) &  &  &  & &   & \\
\noalign{\smallskip}\cline{2-8}\noalign{\smallskip}
     & 20" \& PS1 & 30 & 33 & 0.17 & 0.14 & 0.03 & 0.59\\
     &  \& PS2 (\#3) &  &  & &  &   &\\
\noalign{\smallskip}\hline
\end{tabular}
}
\end{table}

The rate and the signal strength both of the measurement and simulation show in Tab.\,\ref{tab:comparison}, where the measured rate of 20-inch PMT is scaled after a charge cut larger than 40\,p.e. for NNVT PMT and 10\,p.e. for HPK PMT as discussed. The simulation shows a more narrow distribution than measurements where the height of y-axis is artificial normalized, and the measurements have a long tail than the simulation.
As discussed in Sec.\,\ref{2:config}, here the charge in p.e. is calculated with the peak gain, which has a large bias for the NNVT PMT\cite{JUNOPMTgain}. A corrected signal strength\footnote{Here the gain factor 1.6 is used according to \cite{JUNOPMTgain}.} also updated in Tab.\,\ref{tab:comparison} tagged by "Cor.", which is more consistent with simulation. Fig.\,\ref{fig:sim:nnvt:select:com} and Fig.\,\ref{fig:sim:hpk:select:comp} show the charge spectra comparison between simulation and measurements which is normalized to the higher part of each curves, while the NNVT result is scaled with the gain factor already.

\subsection{Comparison of PMT photocathode up and down}
\label{2:comp}

\begin{figure}[!htb]
    \centering
	\begin{subfigure}[c]{0.3\textwidth}
	\centering
	\includegraphics[width=\linewidth]{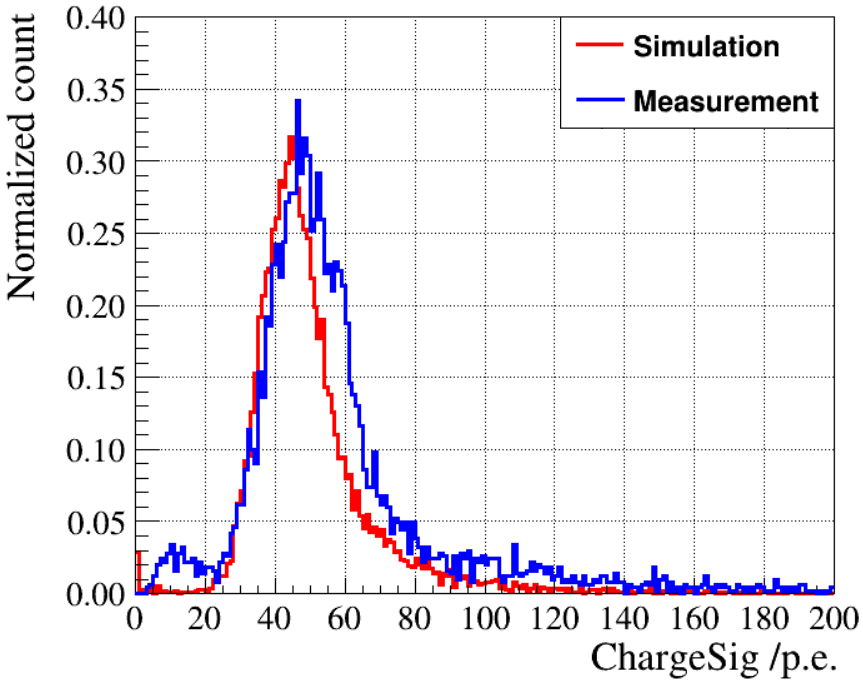}
	\caption{NNVT PMT.}
	\label{fig:sim:nnvt:select:com}
	\end{subfigure}	
	\begin{subfigure}[c]{0.3\textwidth}
	\centering
	\includegraphics[width=\linewidth]{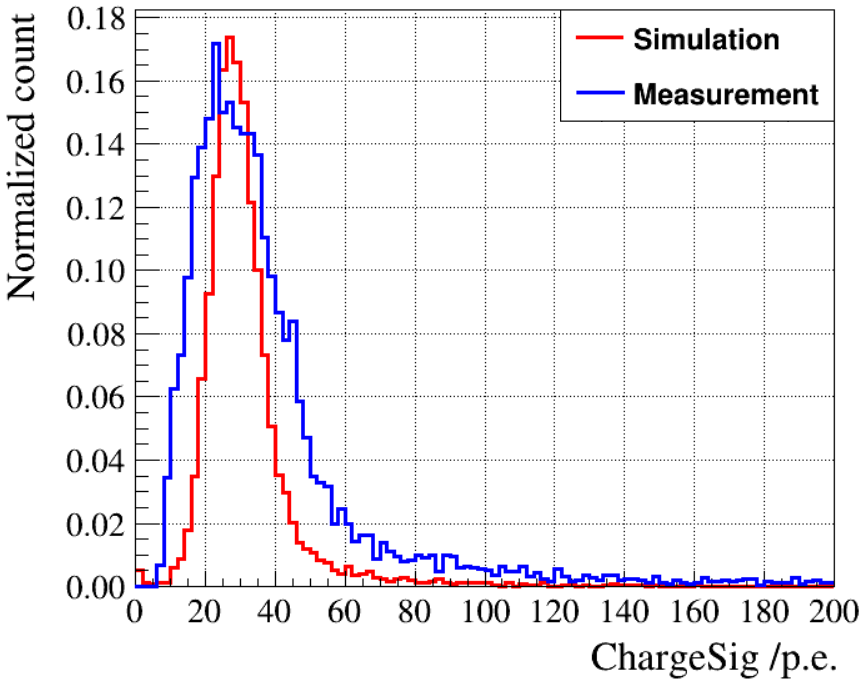}
	\caption{HPK PMT.}
	\label{fig:sim:hpk:select:comp}
	\end{subfigure}
	\begin{subfigure}[c]{0.3\textwidth}
	\centering
	\includegraphics[width=\linewidth]{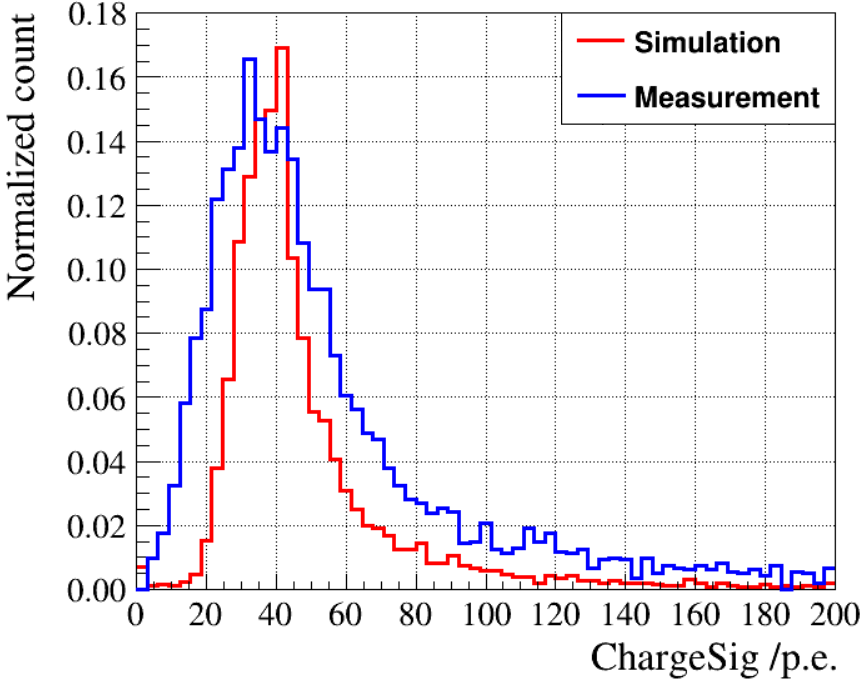}
	\caption{NNVT photocathode down.}
	\label{fig:sim:NNVT:select:comp:photocathode}
	\end{subfigure}
	\caption{Comparison between the measurements and the simulation of 20-inch PMTs.}
	\label{fig:sim:spec:comp}       
\end{figure}

The simulation of the NNVT PMT photocathode towards to down and up is shown in Fig.\,\ref{fig:sim:NNVT:select:comp:photocathode} for NNVT PMT. The reason is similar to that shown in Fig.\ref{fig:sim:hitlocation} that the effect of total reflection from PMT glass to air or vacuum is confirmed. While the collected photon number of PMT photocathode towards to down is less than that of PMT photocathode towards to up. It is around 76\% of the output of photocathode towards to down to photocathode towards to up, which is basically consistent with the measurement.

\section{Summary}
\label{1:summary}

In this paper, we investigated the large pulses of 20-inch PMT dark count, which are mainly from the muons crossing through the PMT glass which generate some large pulses with Cerenkov radiation for both types of dynode and MCP PMTs, and compared their charge responses between the measured and simulated charge spectra. We setup a waveform data taking system to get the muon related PMT large pulse in dark with a setup of muon coincidence system, which is extracted from the peak location of the single photoelectron at different zenith angles, for both PMT types. It is also concluded that the large pulses detected by PMT are able to be interpreted in conjunction with muon hitting PMT at recent world-wide neutrino experiments. It is also confirmed that the signal strength related to muons of photocathode down is smaller than photocathode up, but the difference of muon zenith angles has little effect. It is also confirmed that the charge output of both types of PMTs is linear to the PMT glass thickness and QE coefficient.

\section*{Acknowledgements}

This work was supported by the National Natural Science Foundation of China No. 11875282, the Strategic Priority Research Program of the Chinese Academy of Sciences (Grant No. XDA100102).





\bibliographystyle{unsrtnat}
\bibliography{allcites}   

\begin{thebibliography}{62}
\providecommand{\natexlab}[1]{#1}
\providecommand{\url}[1]{\texttt{#1}}
\expandafter\ifx\csname urlstyle\endcsname\relax
  \providecommand{\doi}[1]{doi: #1}\else
  \providecommand{\doi}{doi: \begingroup \urlstyle{rm}\Url}\fi

\bibitem[Djurcic et~al.(2015)]{JUNOCDR}
Zelimir Djurcic et~al.
\newblock {JUNO Conceptual Design Report}.
\newblock \emph{arXiv e-prints}, 8 2015.

\bibitem[An et~al.(2016)]{JUNOphysics}
F.~An et~al.
\newblock {Neutrino Physics with JUNO}.
\newblock \emph{J. Phys. G}, 43\penalty0 (3):\penalty0 030401, 2016.
\newblock \doi{10.1088/0954-3899/43/3/030401}.

\bibitem[collaboration(2022)]{JUNOdetector}
JUNO collaboration.
\newblock {JUNO} physics and detector.
\newblock \emph{Progress in Particle and Nuclear Physics}, 123:\penalty0
  103927, 2022.
\newblock ISSN 0146-6410.
\newblock \doi{https://doi.org/10.1016/j.ppnp.2021.103927}.
\newblock URL
  \url{https://www.sciencedirect.com/science/article/pii/S0146641021000880}.

\bibitem[Wen et~al.(2019)]{JUNOPMTliangjian}
L.~Wen et~al.
\newblock {A quantitative approach to select PMTs for large detectors}.
\newblock \emph{Nucl. Instrum. Meth. A}, 947:\penalty0 162766, 2019.
\newblock \doi{10.1016/j.nima.2019.162766}.

\bibitem[Qin(2018)]{JUNOPMTinstr}
Z.~Qin.
\newblock {Status of the 20-in. PMT Instrumentation for the JUNO Experiment}.
\newblock In \emph{Proceedings of International Conference on Technology and
  Instrumentation in Particle Physics 2017}, pages 285--293. Springer
  Singapore, 2018.
\newblock ISBN 978-981-13-1316-5.

\bibitem[K.K.(2019)]{HPK-R12860}
Hamamatsu~Photonics K.K.
\newblock {R12860 datasheet}, 2019.
\newblock https://www.hamamatsu.com/jp/en/product/type/R12860/index.html.

\bibitem[Ltd.(2020)]{NNVT-GDB6201-note}
Northern Night Vision~Technology Ltd.
\newblock {Specification for GDB-6201 microchannel plate type photomultiplier
  PMT (in Chinese)}, 2020.
\newblock https://max.book118.com/html/2020/0214/7002125152002115.shtm.

\bibitem[Fukuda et~al.(1999)]{Super-Kamiokande:1998uiq}
Y.~Fukuda et~al.
\newblock {Measurement of the flux and zenith angle distribution of upward
  through going muons by Super-Kamiokande}.
\newblock \emph{Phys. Rev. Lett.}, 82:\penalty0 2644--2648, 1999.
\newblock \doi{10.1103/PhysRevLett.82.2644}.

\bibitem[{Abe K. et al.}(2011)]{PhysRevD.83.052010}
{Abe K. et al.}
\newblock {Solar neutrino results in Super-Kamiokande-III}.
\newblock \emph{Phys. Rev. D}, 83:\penalty0 052010, Mar 2011.
\newblock \doi{10.1103/PhysRevD.83.052010}.
\newblock URL \url{https://link.aps.org/doi/10.1103/PhysRevD.83.052010}.

\bibitem[{Eguchi, K. et al.}({2003})]{PhysRevLett.90.021802}
{Eguchi, K. et al.}
\newblock {First Results from KamLAND: Evidence for Reactor Antineutrino
  Disappearance}.
\newblock \emph{{Phys. Rev. Lett.}}, {90}:\penalty0 {021802}, {Jan} {2003}.
\newblock \doi{"10.1103/PhysRevLett.90.021802"}.

\bibitem[{Ahmad QR, Allen RC, Andersen TC, et al.}(2002)]{SNO}
{Ahmad QR, Allen RC, Andersen TC, et al.}
\newblock {Measurement of day and night neutrino energy spectra at SNO and
  constraints on neutrino mixing parameters}.
\newblock \emph{{Phys Rev Lett.}}, 89\penalty0 (1):\penalty0 011302, 2002.
\newblock \doi{{10.1103/PhysRevLett.89.011302}}.

\bibitem[Lasserre(2012)]{MiniBoone}
Thierry Lasserre.
\newblock The reactor antineutrino anomaly.
\newblock \emph{Journal of Physics Conference Series}, 375:\penalty0 2042--, 07
  2012.
\newblock \doi{10.1088/1742-6596/375/1/042042}.

\bibitem[{Aartsen, M. G. et al.}(2013)]{PhysRevLett.110.131302}
{Aartsen, M. G. et al.}
\newblock {Search for Dark Matter Annihilations in the Sun with the 79-String
  IceCube Detector}.
\newblock \emph{{Phys. Rev. Lett.}}, 110:\penalty0 131302, Mar 2013.
\newblock \doi{10.1103/PhysRevLett.110.131302}.

\bibitem[Y and others"("2012")]{chooz}
"Abe Y and others".
\newblock {Indication of reactor v(e) disappearance in the Double Chooz
  experiment}.
\newblock \emph{"Phys Rev Lett."}, "108"\penalty0 ("13"):\penalty0 "131801",
  "2012".
\newblock \doi{"10.1103/PhysRevLett.108.131801"}.

\bibitem[FP et~al.(2012)]{dayabay}
An~FP et~al.
\newblock {Observation of electron-antineutrino disappearance at Daya Bay}.
\newblock \emph{Phys Rev Lett.}, 108\penalty0 (17):\penalty0 171803, 2012.
\newblock \doi{10.1103/PhysRevLett.108.171803}.

\bibitem[Kim(2013)]{KIM201324}
Soo-Bong Kim.
\newblock {Observation of Reactor Electron Antineutrino Disappearance at RENO}.
\newblock \emph{Nuclear Physics B - Proceedings Supplements}, 235-236:\penalty0
  24--29, 2013.
\newblock ISSN 0920-5632.
\newblock \doi{10.1016/j.nuclphysbps.2013.03.006}.

\bibitem[Barnhill et~al.(2008)]{AugerPMT}
D.~Barnhill et~al.
\newblock {Testing of photomultiplier tubes for use in the surface detector of
  the Pierre Auger Observatory}.
\newblock \emph{Nucl. Instrum. Meth. A}, 591:\penalty0 453--466, 2008.
\newblock \doi{10.1016/j.nima.2008.01.088}.

\bibitem[Ge et~al.(2016)Ge, Zhang, Chen, Cao, Zhang, Wang, and Bi]{GE2016175}
Maomao Ge, Li~Zhang, Yingtao Chen, Zhen Cao, Shoushan Zhang, Chong Wang, and
  Baiyang Bi.
\newblock Photomultiplier tube selection for the wide field of view
  cherenkov/fluorescence telescope array of the large high altitude air shower
  observatory.
\newblock \emph{Nuclear Instruments and Methods in Physics Research Section A},
  819:\penalty0 175--181, 2016.
\newblock \doi{10.1016/j.nima.2016.02.093}.

\bibitem[Ranucci et~al.(1993)]{BorexinoPMT}
G.~Ranucci et~al.
\newblock {Characterization and magnetic shielding of the large cathode area
  PMTs used for the light detection system of the prototype of the solar
  neutrino experiment Borexino}.
\newblock \emph{Nucl. Instrum. Meth. A}, 337:\penalty0 211--220, 1993.
\newblock \doi{10.1016/0168-9002(93)91156-H}.

\bibitem[Liu(2008)]{DayabayPMT}
D.~Liu.
\newblock {PMT evaluation for the Daya Bay neutrino experiment}.
\newblock In \emph{{2008 IEEE Nuclear Science Symposium and Medical Imaging
  Conference and 16th International Workshop on Room-Temperature Semiconductor
  X-Ray and Gamma-Ray Detectors}}, pages 3133--3139, 2008.
\newblock \doi{10.1109/NSSMIC.2008.4775017}.

\bibitem[Baldini et~al.(1996)]{ChoozPMT}
A.~Baldini et~al.
\newblock {The photomultiplier test facility for the reactor neutrino
  oscillation experiment CHOOZ and the measurements of 250 8-in. EMI 9356KA B53
  photomultipliers}.
\newblock \emph{Nucl. Instrum. Meth. A}, 372\penalty0 (1):\penalty0 207--221,
  1996.
\newblock \doi{https://doi.org/10.1016/0168-9002(95)01236-2}.

\bibitem[Bronner et~al.(2020)]{HKPMT}
C.~Bronner et~al.
\newblock {Development and performance of the 20'' PMT for Hyper-Kamiokande}.
\newblock \emph{J. Phys. Conf. Ser.}, 1468\penalty0 (1):\penalty0 012237, 2020.
\newblock \doi{10.1088/1742-6596/1468/1/012237}.

\bibitem[Cao et~al.(2021)]{JUNO3inchPMT}
C.~Cao et~al.
\newblock {Mass production and characterization of 3-inch PMTs for the JUNO
  experiment}.
\newblock \emph{Nucl. Instrum. Meth. A}, 1005:\penalty0 165347, 2021.
\newblock \doi{10.1016/j.nima.2021.165347}.

\bibitem[Mollo(2016)]{KM3NeTPMT}
C.~M. Mollo.
\newblock {Development and performances of a high statistics PMT test
  facility}.
\newblock \emph{EPJ Web Conf.}, 116:\penalty0 06010, 2016.
\newblock \doi{10.1051/epjconf/201611606010}.

\bibitem[Yang et~al.(2020{\natexlab{a}})]{JUNOPMTflasher}
A.~Yang et~al.
\newblock {Study and removal of the flash from the HV divider of the 20-inch
  PMT for JUNO}.
\newblock \emph{JINST}, 15\penalty0 (04):\penalty0 T04006, 2020{\natexlab{a}}.
\newblock \doi{10.1088/1748-0221/15/04/T04006}.

\bibitem[Qian et~al.(2019)]{MCPPMT2018}
S.~Qian et~al.
\newblock {The improvement of 20'' MCP-PMT for neutrino detection}.
\newblock \emph{PoS}, ICHEP2018:\penalty0 662, 2019.
\newblock \doi{10.22323/1.340.0662}.

\bibitem[Wang et~al.(2012)]{YWang_newMCP}
Y.~Wang et~al.
\newblock {A new design of large area MCP-PMT for the next generation neutrino
  experiment}.
\newblock \emph{Nucl. Instrum. Meth. A}, 695:\penalty0 113--117, 2012.
\newblock \doi{10.1016/j.nima.2011.12.085}.

\bibitem[Yin et~al.(2018)]{wavesamplingPMT}
S.~Yin et~al.
\newblock {A novel PMT test system based on waveform sampling}.
\newblock \emph{JINST}, 13\penalty0 (01):\penalty0 T01005, 2018.
\newblock \doi{10.1088/1748-0221/13/01/T01005}.

\bibitem[Zhang et~al.(2019{\natexlab{a}})]{waveAnalysisHaiqiong}
H.~Q. Zhang et~al.
\newblock {Comparison on PMT Waveform Reconstructions with JUNO Prototype}.
\newblock \emph{JINST}, 14\penalty0 (08):\penalty0 T08002, 2019{\natexlab{a}}.
\newblock \doi{10.1088/1748-0221/14/08/T08002}.

\bibitem[et~al.(2016)]{Abe_2016}
Y.~Abe et~al.
\newblock Characterization of the spontaneous light emission of the {PMTs} used
  in the double chooz experiment.
\newblock \emph{Journal of Instrumentation}, 11\penalty0 (08):\penalty0
  P08001--P08001, aug 2016.
\newblock \doi{10.1088/1748-0221/11/08/p08001}.

\bibitem[Dwyer(2013)]{DWYER201330}
D.A. Dwyer.
\newblock Improved measurement of electron-antineutrino disappearance at daya
  bay.
\newblock \emph{Nuclear Physics B}, 235-236:\penalty0 30--32, 2013.
\newblock \doi{10.1016/j.nuclphysbps.2013.03.007}.

\bibitem[van Eijk et~al.(2020)van Eijk, Schneider, and
  Unland]{IceCube-inproceedings}
D.~van Eijk, J.~Schneider, and M.~Unland.
\newblock {Characterisation of Two PMT Models for the IceCube Upgrade mDOM}.
\newblock \emph{PoS}, ICRC2019:\penalty0 1022, 2020.
\newblock \doi{10.22323/1.358.1022}.

\bibitem[Jang(2014)]{JANG2014145}
J.S. Jang.
\newblock A precise measurement of reactor antineutrino at reno.
\newblock \emph{Nuclear Data Sheets}, 120:\penalty0 145--148, 2014.
\newblock ISSN 0090-3752.
\newblock \doi{10.1016/j.nds.2014.07.030}.

\bibitem[Yang et~al.(2020{\natexlab{b}})Yang, Qin, Wang, Chen, Wei, Luo, Xu,
  Heng, and Ouyang]{Yang_2020}
A.~Yang, Z.~Qin, Z.~Wang, H.~Chen, W.~Wei, F.~Luo, M.~Xu, Y.~Heng, and
  Q.~Ouyang.
\newblock Study and removal of the flash from the {HV} divider of the 20-inch
  {PMT} for {JUNO}.
\newblock \emph{Journal of Instrumentation}, 15\penalty0 (04):\penalty0
  T04006--T04006, apr 2020{\natexlab{b}}.
\newblock \doi{10.1088/1748-0221/15/04/t04006}.

\bibitem[Qian et~al.(2020)Qian, Gao, Ma, Zhu, Chen, and Chen]{Qian_2020}
Sen Qian, Feng Gao, Lishuang Ma, Yao Zhu, Shanhong Chen, and Pengyu Chen.
\newblock The study on the 20 inch {PMT} flasher signal.
\newblock \emph{Journal of Instrumentation}, 15\penalty0 (06):\penalty0
  T06008--T06008, 2020.
\newblock \doi{10.1088/1748-0221/15/06/t06008}.

\bibitem["W.~Raposo("2007")]{PMTmuon2007}
L.~Villasenor" "W.~Raposo, M.~Vaz.
\newblock "measurements of signals from muons crossing the hamamatsu r5912 pmt
  enclosure vertically and horizontally".
\newblock \emph{online}, "2007".
\newblock "AngraNote 004-2007".

\bibitem[Bayat et~al.(2014)Bayat, Doust-Mohammadi, Ghorbani, Ghal-Eh, and
  Mohammadi]{BAYAT20141}
E.~Bayat, V.~Doust-Mohammadi, P.~Ghorbani, N.~Ghal-Eh, and R.~Mohammadi.
\newblock Scintillation of xp2020 pmt glass window.
\newblock \emph{Radiation Physics and Chemistry}, 102:\penalty0 1--4, 2014.
\newblock ISSN 0969-806X.
\newblock \doi{https://doi.org/10.1016/j.radphyschem.2014.04.009}.
\newblock URL
  \url{https://www.sciencedirect.com/science/article/pii/S0969806X14001285}.

\bibitem[{Hyper-Kamiokande Proto-Collaboration} et~al.(2018){Hyper-Kamiokande
  Proto-Collaboration}, K, et~al.]{Hyper-Kamiokande:2018ofw}
{Hyper-Kamiokande Proto-Collaboration}, {Abe} K, et~al.
\newblock {Hyper-Kamiokande Design Report}.
\newblock \emph{arXiv e-prints}, art. arXiv:1805.04163, May 2018.

\bibitem[Abe(2015)]{HK10.1093/ptep/ptv061}
K.~et~al. Abe.
\newblock {Physics potential of a long-baseline neutrino oscillation experiment
  using a J-PARC neutrino beam and Hyper-Kamiokande}.
\newblock \emph{Progress of Theoretical and Experimental Physics},
  2015\penalty0 (5), 05 2015.
\newblock \doi{10.1093/ptep/ptv061}.
\newblock URL \url{https://doi.org/10.1093/ptep/ptv061}.
\newblock 053C02.

\bibitem[TT112(2021)]{TT112}
Shanghai TT112.
\newblock {thickness meter TT112}, 2021.
\newblock http://www.jsbyb.com/wujiu-Products-4070951/.

\bibitem[et~al.(2015)]{R5912-CHOW201525}
Ken~Chow et~al.
\newblock Waterproofed photomultiplier tube assemblies for the daya bay reactor
  neutrino experiment.
\newblock \emph{Nuclear Instruments and Methods in Physics Research Section A},
  794:\penalty0 25--32, 2015.
\newblock \doi{10.1016/j.nima.2015.05.002}.

\bibitem[Photonis(1998)]{XP2020}
Photonis.
\newblock Standard very fast, 12-stage, 51 mm (2") round tube, 1998.
\newblock URL \url{https://irfu.cea.fr/dphn/Tp/xp2020.pdf}.

\bibitem[{CAEN SpA}(2021{\natexlab{a}})]{CAEN-DT5751}
{CAEN SpA}.
\newblock {DT5751}, 2/4 channel 10 bit 2/1 gs/s digitizer, 2021{\natexlab{a}}.
\newblock https://www.caen.it/products/dt5751/.

\bibitem[iseg Spezialelektronik GmbH~company(2021)]{iseg-SHR}
iseg Spezialelektronik GmbH~company.
\newblock {SHR}, switchable high end high precision ac/dc desktop hv
  source-measure-unit, 2021.
\newblock https://iseg-hv.com/en/products/detail/SHR.

\bibitem[{CAEN SpA}(2021{\natexlab{b}})]{CAEN-N625}
{CAEN SpA}.
\newblock {N625}, quad linear fan-in fan-out, 2021{\natexlab{b}}.
\newblock https://www.caen.it/products/n625/.

\bibitem[{CAEN SpA}(2021{\natexlab{c}})]{CAEN-N979}
{CAEN SpA}.
\newblock {N979}, 16 channel fast amplifier, 2021{\natexlab{c}}.
\newblock https://www.caen.it/products/n979/.

\bibitem[{CAEN SpA}(2021{\natexlab{d}})]{CAEN-N845}
{CAEN SpA}.
\newblock {N845}, 16 channel low threshold discriminator, 2021{\natexlab{d}}.
\newblock https://www.caen.it/products/n845/.

\bibitem[{CAEN SpA}(2021{\natexlab{e}})]{CAEN-N1145}
{CAEN SpA}.
\newblock {N1145}, quad scaler and preset counter/timer, 2021{\natexlab{e}}.
\newblock https://www.caen.it/products/n1145/.

\bibitem[{CAEN SpA}(2021{\natexlab{f}})]{CAEN-N455}
{CAEN SpA}.
\newblock {N455}, quad coincidence logic unit, 2021{\natexlab{f}}.
\newblock https://www.caen.it/products/n455/.

\bibitem[Luo et~al.(2018)]{JUNOPMTsignalopt}
F.~Luo et~al.
\newblock {Signal Optimization with HV divider of MCP-PMT for JUNO}.
\newblock \emph{Springer Proc. Phys.}, 213:\penalty0 309--314, 2018.
\newblock \doi{10.1007/978-981-13-1316-5\_58}.

\bibitem[Luo et~al.(2016)]{JUNOPMTsignalover}
F.~Luo et~al.
\newblock {PMT overshoot study for the JUNO prototype detector}.
\newblock \emph{Chin. Phys. C}, 40\penalty0 (9):\penalty0 096002, 2016.
\newblock \doi{10.1088/1674-1137/40/9/096002}.

\bibitem[Polyakov(2013)]{POLYAKOV201369}
Sergey~V. Polyakov.
\newblock Chapter 3 - photomultiplier tubes.
\newblock In Alan~Migdall et~al., editor, \emph{Single-Photon Generation and
  Detection}, volume~45 of \emph{Experimental Methods in the Physical
  Sciences}, pages 69--82. Academic Press, 2013.
\newblock \doi{10.1016/B978-0-12-387695-9.00003-2}.

\bibitem[Bellamy et~al.(1994)]{PMTgainmodel1994}
E.~H. Bellamy et~al.
\newblock {Absolute calibration and monitoring of a spectrometric channel using
  a photomultiplier}.
\newblock \emph{Nucl. Instrum. Meth. A}, 339:\penalty0 468--476, 1994.
\newblock \doi{10.1016/0168-9002(94)90183-X}.

\bibitem[et~al.(2019)]{Luo_2019}
F.~Luo et~al.
\newblock A study of the new hemispherical 9-inch {PMT}.
\newblock \emph{Journal of Instrumentation}, 14\penalty0 (02):\penalty0
  T02004--T02004, feb 2019.
\newblock \doi{10.1088/1748-0221/14/02/t02004}.
\newblock URL \url{https://doi.org/10.1088/1748-0221/14/02/t02004}.

\bibitem[Saldanha et~al.(2017)]{PMTgainmodel2017}
R.~Saldanha et~al.
\newblock {Model Independent Approach to the Single Photoelectron Calibration
  of Photomultiplier Tubes}.
\newblock \emph{Nucl. Instrum. Meth. A}, 863:\penalty0 35--46, 2017.
\newblock \doi{10.1016/j.nima.2017.02.086}.

\bibitem[Zhang et~al.(2021)]{JUNOPMTgain}
H.~Q. Zhang et~al.
\newblock {Gain and charge response of 20'' MCP and dynode PMTs}.
\newblock \emph{JINST}, 16\penalty0 (08):\penalty0 T08009, 2021.
\newblock \doi{10.1088/1748-0221/16/08/T08009}.

\bibitem[{Hamamatsu Photonics K.K}(2007)]{HamManual}
{Hamamatsu Photonics K.K}.
\newblock \emph{{Photomultiplier Tubes - Basics and Applications}}.
\newblock Hamamatsu Photonics K.K., {3rd} edition, 2007.
\newblock https://www.hamamatsu.com/resources/pdf/etd/PMT\_handbook\_v3aE.pdf.

\bibitem[Coordinators(2021)]{geant4}
Geant4 Working Groups~\& Coordinators.
\newblock {Geant4}, 2021.
\newblock https://geant4.web.cern.ch/.

\bibitem[et~al.(2017)]{Lin_2017}
Tao~Lin et~al.
\newblock The application of {SNiPER} to the {JUNO} simulation.
\newblock \emph{Journal of Physics: Conference Series}, 898:\penalty0 042029,
  oct 2017.
\newblock \doi{10.1088/1742-6596/898/4/042029}.
\newblock URL \url{https://doi.org/10.1088/1742-6596/898/4/042029}.

\bibitem[Guan et~al.(2015)Guan, chung Chu, Cao, Luk, and gen
  Yang]{Muonflux-Guan2015APO}
Mengyun Guan, Ming chung Chu, Jun Cao, K.~B. Luk, and Chang gen Yang.
\newblock A parametrization of the cosmic-ray muon flux at sea-level.
\newblock \emph{arXiv: High Energy Physics - Experiment}, 2015.

\bibitem[Zhang et~al.(2019{\natexlab{b}})]{PMTrelativeCE}
H.~Zhang et~al.
\newblock {Study on relative collection efficiency of PMTs with spotlight}.
\newblock \emph{Radiat. Detect. Technol. Methods}, 3:\penalty0 20,
  2019{\natexlab{b}}.
\newblock \doi{10.1007/s41605-019-0099-x}.

\bibitem[Lei et~al.(2016)]{largePMTlei}
X.-C. Lei et~al.
\newblock {Evaluation of new large area PMT with high quantum efficiency}.
\newblock \emph{Chin. Phys. C}, 40\penalty0 (2):\penalty0 026002, 2016.
\newblock \doi{10.1088/1674-1137/40/2/026002}.

\end{thebibliography}








\end{document}